\documentclass[lettersize,journal]{IEEEtran}
\usepackage{amsmath,amsfonts}
\usepackage{algorithmic}
\usepackage{algorithm}
\usepackage{array}
\usepackage[caption=false,font=normalsize,labelfont=sf,textfont=sf]{subfig}
\usepackage{textcomp}
\usepackage{stfloats}
\usepackage{url}
\usepackage{verbatim}
\usepackage{graphicx}

\usepackage{cite}
\usepackage[normalem]{ulem}
\newcommand{\stkout}[1]{\ifmmode\text{\sout{\ensuremath{#1}}}\else\sout{#1}\fi}

\usepackage[acronym,shortcuts]{glossaries}
\newacronym[firstplural=spatio-temporal covariance matrices (STCMs)]{STCM}{STCM}{spatio-temporal covariance matrix}
\newacronym{BLCMP}{BLCMP}{binaural linearly constrained minimum power}
\newacronym{RIR}{RIR}{room impulse response}
\newacronym{LCMP}{LCMP}{linearly constrained minimum power}
\newacronym{wBLCMP}{wBLCMP}{weighted binaural linearly constrained minimum power}
\newacronym{STFT}{STFT}{short-time Fourier transform}
\newacronym{CTF}{CTF}{convolutive transfer function}
\newacronym{MTF}{MTF}{multiplicative transfer function}
\newacronym{RTF}{RTF}{relative transfer function}
\newacronym{MCLP}{MCLP}{multi channel linear prediction}
\newacronym{MPDR}{MPDR}{minimum power distortionless response}
\newacronym{MVDR}{MVDR}{minimum variance distortionless response}
\newacronym{LCMV}{LCMV}{linearly constrained minimum variance}
\newacronym{WPD}{WPD}{weighted power minimization distortionless response}
\newacronym{WPE}{WPE}{weighted prediction error}
\newacronym{TVG}{TVG}{time-varying complex circular Gaussian}
\newacronym{MISO}{MISO}{multiple-input single-output}
\newacronym{MIMO}{MIMO}{multiple-input multiple-output}
\newacronym{SPP}{SPP}{speech presence probability}
\newacronym{PESQ}{PESQ}{perceptual evaluation of speech quality}
\newacronym{FWSSNR}{FWSSNR}{frequency-weighted segmental signal-to-noise ratio}
\newacronym{CW}{CW}{covariance whitening}
\newacronym{CS}{CS}{covariance subtraction}
\newacronym{VAD}{VAD}{voice activity detection}
\newacronym{SAD}{SAD}{source activity detection}
\newacronym{UCB}{UCB}{unified convolutional beamformer}
\newacronym{IRLS}{IRLS}{iteratively reweighted least squares}
\newacronym{MFMVDR}{MFMVDR}{multi-frame minimum variance distortionless response}
\newacronym{BMFMVDR}{BMFMVDR}{binaural MFMVDR}
\newacronym{TCN}{TCN}{temporal convolutional network}
\newacronym{DNN}{DNN}{deep neural network}
\newacronym{SNR}{SNR}{signal-to-noise ratio}
\newacronym{DRC}{DRC}{dynamic range compression}
\newacronym{HGR}{HGR}{half-gain rule}
\newacronym{CEC1}{CEC1}{Clarity Enhancement Challenge}
\newacronym{MBDRC}{MBDRC}{multi-band dynamic range compressor}
\newacronym{MBSTOI}{MBSTOI}{modified binaural short-term objective intelligibility}
\newacronym{SD-SDR}{SD-SDR}{scale-dependent signal-to-distortion ratio}
\newacronym{IFC}{IFC}{inter-frame correlation}
\newacronym{BTE}{BTE}{behind-the-ear}
\newacronym{SIR}{SIR}{signal-to-interferer ratio}
\newacronym{SINR}{SINR}{signal-to-interferer-and-noise ratio}
\newacronym{SRR}{SRR}{signal-to-reverberation ratio}
\newacronym{DUR}{DUR}{desired-to-undesired ratio}
\newacronym{wMPDR}{wMPDR}{weighted \gls{MPDR}}
\newacronym{SVD}{SVD}{singular value decomposition}
\newacronym{LS}{LS}{least-squares}
\newacronym{CBW}{CBW}{covariance blocking and whitening}
\newacronym{SRI}{SRI}{successive \gls{RTF} identification}
\newacronym{BOP-SRI}{BOP-SRI}{blind oblique projection \gls{SRI}}
\newacronym{BOP}{BOP}{blind oblique projection}
\newacronym{EBOP}{EBOP}{extended blind oblique projection}
\newacronym{CWu}{CWu}{\gls{CW} with the undesired covariance matrix}
\newacronym[\glslongpluralkey=power spectral densities]{PSD}{PSD}{power spectral density}
\newacronym{HA}{HA}{Hermitian angle}
\newacronym{MHA}{MHA}{mean Hermitian angle}
\newacronym{WMHA}{WMHA}{weighted mean Hermitian angle}
\newacronym{GMSC}{GMSC}{generalized magnitude-squared coherence}
\newacronym{LTASS}{LTASS}{long-term average speech spectrum}
\newacronym{EM}{EM}{expectation maximization}
\newacronym{CFA}{CFA}{confirmatory factor analysis}
\newacronym{DOA}{DOA}{direction-of-arrival}
\newacronym{BOPO}{BOPO}{BOP using orthogonal additional vectors}
\newacronym{BOP-S}{BOP-S}{BOP with noise subtraction}
\newacronym{BOP-W}{BOP-W}{BOP with noise whitening}
\newacronym{BOPO-S}{BOPO-S}{BOPO with noise subtraction}
\newacronym{BOPO-W}{BOPO-W}{BOPO with noise whitening}

\usepackage[hidelinks]{hyperref}
\usepackage{cleveref}
\crefname{equation}{}{}
\crefformat{appsec}{Appendix #2#1#3} 
\usepackage{amssymb,mathtools,nicefrac,bm}

\DeclareMathOperator*{\argmin}{argmin}
\DeclarePairedDelimiter\abs{\lvert}{\rvert}%
\DeclarePairedDelimiter\norm{\lVert}{\rVert}%
\makeatletter
\let\oldabs\abs
\def\abs{\@ifstar{\oldabs}{\oldabs*}}
\let\oldnorm\norm
\def\norm{\@ifstar{\oldnorm}{\oldnorm*}}
\makeatother

\usepackage{siunitx}
\usepackage{microtype}
\usepackage{multirow}
\usepackage{booktabs}

\begin{document}

\title{Closed-Form Successive Relative Transfer Function Vector Estimation
based on Blind Oblique Projection Incorporating Noise Whitening}

\author{Henri Gode 
    and Simon Doclo,~\IEEEmembership{Senior Member,~IEEE}
    \thanks{The authors are with the Department of Medical Physics and Acoustics and the Cluster of Excellence Hearing4all, University of Oldenburg, Germany (e-mail: henri.gode@uni-oldenburg.de; simon.doclo@uni-oldenburg.de).}
    
    
}

\markboth{Journal of \LaTeX\ Class Files,~Vol.~14, No.~8, August~2021}%
{Shell \MakeLowercase{\textit{et al.}}: A Sample Article Using IEEEtran.cls for IEEE Journals}


\maketitle





This work has been submitted to the IEEE for possible publication. Copyright may be transferred without notice, after which this version may no longer be accessible.
\newline

\begin{abstract}
    \Glspl{RTF} of sound sources play a crucial role in beamforming, enabling effective noise and interference suppression.
    This paper addresses the challenge of online estimating the \gls{RTF} vectors of multiple sound sources
    in noisy and reverberant environments, for the specific scenario where sources activate successively.
    While the \gls{RTF} vector of the first source can be estimated straightforwardly,
    the main challenge arises in estimating the \gls{RTF} vectors of subsequent sources
    during segments where multiple sources are simultaneously active.
    The \gls{BOP} method has been proposed to estimate the \gls{RTF} vector of a newly activating source
    by optimally blocking this source.
    However, this method faces several limitations:
    high computational complexity due to its reliance on iterative gradient descent optimization,
    the introduction of random additional vectors, which can negatively impact performance,
    and the assumption of high \gls{SNR}.
    To overcome these limitations, in this paper we propose three extensions to the \gls{BOP} method.
    First, we derive a closed-form solution for optimizing the \gls{BOP} cost function,
    significantly reducing computational complexity.
    Second, we introduce orthogonal additional vectors instead of random vectors,
    enhancing \gls{RTF} vector estimation accuracy.
    Third, we incorporate noise handling techniques inspired by covariance subtraction and whitening,
    increasing robustness in low \gls{SNR} conditions.
    To provide a frame-by-frame estimate of the source activity pattern,
    required by both the conventional \gls{BOP} method and the proposed method,
    we propose a spatial-coherence-based online source counting method.
    Simulations are performed with real-world reverberant noisy recordings featuring 3 successively activating speakers,
    with and without a-priori knowledge of the source activity pattern.
    Simulation results demonstrate that the proposed method outperforms the conventional \gls{BOP} method
    in terms of computational efficiency, \gls{RTF} vector estimation accuracy,
    and signal-to-interferer-and-noise ratio improvement
    when applied in a linearly constrained minimum variance beamformer.
\end{abstract}

\glsresetall

\section{Introduction}

\IEEEPARstart{I}{n} many hands-free speech communication applications, such as hearing aids, mobile phones, and smart speakers,
interfering sounds and ambient noise may degrade the recorded microphone signals~\cite{beutelmann_prediction_2006}.
In such applications, online speech enhancement methods, which rely only on past information,
are required to improve the quality and intelligibility of a target speaker.
When multiple microphones are available, beamforming is a widely used technique
to enhance a target speaker while suppressing interfering sources and noise~\cite{veen_beamforming_1988,doclo_multichannel_2015,gannot_consolidated_2017}.
Commonly used beamformers are the \gls{MVDR} beamformer and the \gls{LCMV} beamformer \cite{veen_beamforming_1988}.
Besides an estimate of the noise covariance matrix,
these beamformers require an estimate of the \gls{RTF} vector of the target source
and possibly also the \gls{RTF} vectors of the interfering sources.
The \gls{RTF} vector relates the acoustic transfer functions between a source and all microphones
to a reference microphone,
and plays an important role not only in beamforming, but also in, e.g.,
source localization \cite{braun2015narrowband,farmani2018bias,grumiaux2022survey,fejgin2022coherence}
and joint noise reduction and dereverberation \cite{nakatani_unified_2019,gode2022adaptive}.

Over the last decades, several methods have been proposed to estimate the \gls{RTF} vector of a single source in a noisy environment, e.g.,
based on (weighted) least-squares~\cite{Gannot2001, cohen_relativ_2004, tammen2019joint},
using maximum likelihood estimation~\cite{li2023joint}, exploiting frequency correlations~\cite{bologni2025wideband},
using subspace decomposition methods such as \gls{CS} and
\gls{CW}~\cite{warsitz2007blind, markovich_multichannel_2009, serizel_low-rank_2014, varzandeh2017iterative, markovich-golan_performance_2018},
or using manifold learning~\cite{brendel2022manifold,levi2024peerrtf}.
The \gls{CS} method estimates the \gls{RTF} vector by computing the principal eigenvector of the noisy covariance matrix after subtracting an estimate of the noise covariance matrix.
On the other hand, the \gls{CW} method estimates the \gls{RTF} vector by de-whitening the principal eigenvector of the whitened noisy covariance matrix,
where an estimate of the noise covariance matrix is used for the whitening operation.
A performance comparison in~\cite{markovich-golan_performance_2018} demonstrated
that the \gls{CW} method achieves superior results compared to the \gls{CS} method.

In contrast to estimating the \gls{RTF} vector of a single source, estimating the \gls{RTF} vectors of multiple simultaneously active sources is more challenging.
It has been shown in~\cite{markovich_multichannel_2009} that
in a multi-source scenario the \gls{CS} and \gls{CW} methods can only estimate the subspace spanning the \gls{RTF} vectors of all sources,
instead of the individual \gls{RTF} vectors.
A generalization of the \gls{RTF} concept to multiple sources was proposed in~\cite{deleforge2015generalized},
where a Plücker spectrogram transform was introduced to define a joint multi-source RTF representation.
This generalized \gls{RTF} retains key properties of single-source RTFs, such as invariance to source signals and dependency only on spatial properties,
but it does not estimate the individual RTF vectors.
Several methods have been proposed which aim at estimating the \gls{RTF} vectors of multiple sources.
In~\cite{markovich2016combined}, an estimate of the \gls{RTF} vector for each source is obtained
using the spatial filters from the TRINICON blind source separation framework~\cite{buchner2004trinicon}.
For this task, TRINICON needs to be geometrically constrained for each source
using an estimate of the \gls{DOA} of that source and assuming the microphone array geometry to be known.
In~\cite{schwartz2017two}, two \gls{EM} algorithms
are derived assuming W-disjoint orthogonality of the sources in the time-frequency domain,
where the \gls{RTF} vectors of multiple sources are estimated alongside other acoustic parameters.
However, the initialization of the \gls{EM} algorithms is challenging and relies on long-term covariance matrices,
the iterative \gls{EM} optimization has a high computational complexity,
and post-processing is required to resolve permutation ambiguity of the sources in each frequency bin.
In~\cite{koutrouvelis2019robust}, the \gls{RTF} vectors of multiple sources are estimated alongside other acoustic parameters
by combining \gls{CFA} and non-orthogonal joint diagonalization and utilizing linear inequality constraints on the parameters to increase robustness.
Unlike the \gls{EM}-based methods, this method does not assume single-source activity per time-frequency bin,
but solving the constrained optimization problem is computationally complex and post-processing is also required to resolve permutation ambiguity.
In~\cite{li2025multimicrophone},
a joint diagonalization method utilizing a Jacobi-like algorithm is introduced to estimate the \gls{RTF} vectors of multiple sources,
reducing computational complexity compared to the \gls{CFA} method while maintaining estimation accuracy, but still requiring post-processing to resolve permutation ambiguity.
In~\cite{dietzen2020square}, a recursive \gls{RTF} vector update method is introduced
based on an orthogonal Procrustes problem aiming at estimating the \gls{RTF} vectors and \glspl{PSD} of all sources.
However, this method requires a good initialization of the \gls{RTF} vectors and assumes the microphone array geometry to be known.

\begin{figure}[t]
    \centering
    \includegraphics[width=1\columnwidth]{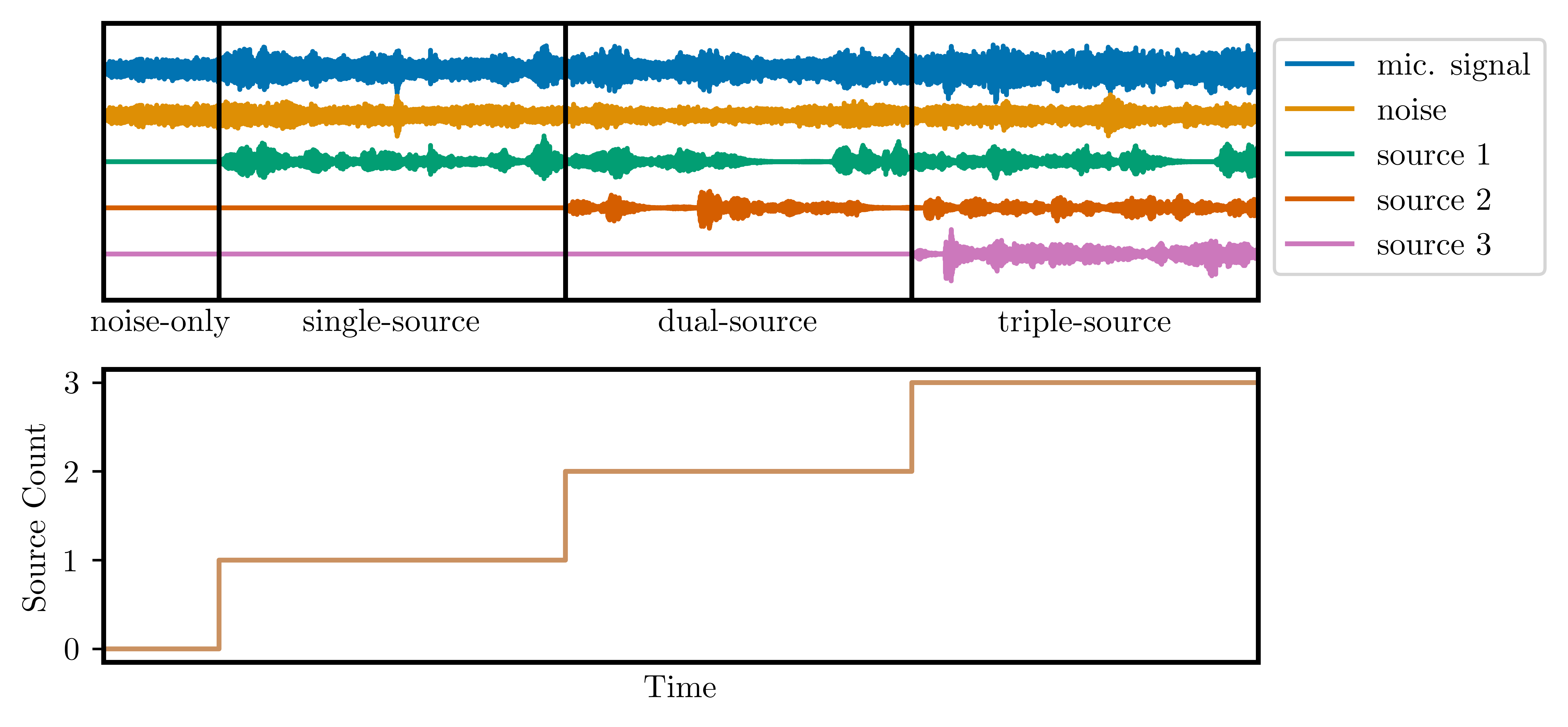}
    \caption{Exemplary scenario with three successively activating sources.
        Upper: Waveforms of one microphone signal and its speech and noise components for four different time segments.
        Lower: Increasing source count over time due to successive activation of sources.}
    \label{fig:segments}
\end{figure}

In this paper, we consider a specific scenario where sources activate successively (see~\Cref{fig:segments} for an exemplary scenario with three sources).
In this scenario, the \gls{RTF} vector of the first source can be estimated
straightforwardly in the single-source segment,
e.g., using the \gls{CW} method with an estimate of the noise covariance matrix obtained from the noise-only segment.
The focus of this paper is on online estimation of the \gls{RTF} vectors of the subsequent sources
(i.e., the second source in the dual-source segment and the third source in the triple-source segment),
using estimates of the noise covariance matrix and the \gls{RTF} vectors of already active sources.
Several methods have already been proposed for successive \gls{RTF} vector estimation.
The covariance blocking and whitening method proposed in~\cite{gode2023covariance}
for a dual-source scenario estimates the \gls{RTF} vector of the second activating source
by solving a set of non-linear equations.
The \gls{BOP} method proposed in~\cite{cherkassky_successive_2020} estimates the \gls{RTF} vector of a newly activating source
by determining the oblique projection operator which optimally blocks this source while keeping the already active sources distortionless.
Despite its accuracy in scenarios with high \gls{SNR}, the conventional \gls{BOP} method exhibits several limitations.
First, the \gls{BOP} method relies on an iterative gradient descent method to minimize its cost function,
possibly converging to a local minimum and resulting in high computational complexity.
Second, to avoid plateaus in its cost function, the \gls{BOP} method employs random additional vectors.
However, when the rank-$1$ approximation of sources is violated in practice,
these random vectors can negatively impact the \gls{RTF} vector estimation accuracy.
Third, the \gls{BOP} method is designed for high \glspl{SNR}, limiting its performance at low \glspl{SNR}.
As for all successive \gls{RTF} vector estimation methods,
it should be realized that the \gls{BOP} method requires knowledge about the source activity pattern,
i.e., when sources become active and inactive.\footnote{In this paper, we will only consider the case where sources enter the scene (source activation).}
To determine the source activity pattern,
several methods have been proposed for counting the number of active sources within a given signal segment.
Single-microphone methods often rely on \glspl{DNN},
as in~\cite{stoter2018countnet,Chetupalli_speakercount_2023},
while common multi-microphone methods perform clustering on spatial features (e.g., narrowband \gls{DOA} estimates) per time-frequency bin,
assuming knowledge of the microphone array geometry~\cite{pavlidi2013real,wang2016iterative,hafezi2019spatial}.
In~\cite{Laufer_source_counting_2018}, the number of sources was based on Blind Oblique Projection (BOP) and the eigenvalue distribution of a spatial correlation matrix.
This method was extended in~\cite{hsu2023learning} by incorporating spectral whitening
to construct a spatial coherence matrix instead of the spatial correlation matrix.
However, it should be realized that all aforementioned methods either estimate the number of sources over relatively long signal segments (1.6-20 seconds),
which is too slow for reliably detecting newly activating sources for the considered application scenarios,
or require knowledge of the microphone array geometry, which is not always available in practice.

To address the limitations of the conventional \gls{BOP} method~\cite{cherkassky_successive_2020}, in this paper we propose three extensions.
First, we derive a closed-form solution to the BOP optimization problem, enhancing robustness and significantly reducing computational complexity.
Second, we propose to utilize orthogonal additional vectors instead of random vectors,
reducing estimation errors caused by model mismatch.
Third, we incorporate noise handling techniques inspired by the \gls{CS} and \gls{CW} methods
for single-source \gls{RTF} vector estimation, making the BOP method suitable for low \gls{SNR} conditions.
Instead of assuming a-priori knowledge of the source activity pattern,
we also propose a spatial-coherence-based online source counting method
to provide a frame-by-frame estimate of the source activity pattern,
allowing for a more realistic application
of the proposed multi-source \gls{RTF} vector estimation method.
For three successively activating sources in a reverberant noisy environment, simulation results for a binaural hearing aid setup with six microphones demonstrate
that the proposed method featuring all three extensions outperforms the conventional \gls{BOP} method
for several source positions and \glspl{SNR}.
The improvements are evident with and without a-priori knowledge
of the source activity pattern in terms of computational complexity,
\gls{RTF} vector estimation accuracy,
and signal-to-interferer-and-noise ratio improvement
when the estimated \gls{RTF} vectors are used in an \gls{LCMV} beamformer.

The remainder of this paper is organized as follows.
\Cref{sec:signal_model} describes the signal model used for \gls{RTF} vector estimation of successively activating sources.
\Cref{sec:LCMV} briefly reviews the \gls{LCMV} beamformer,
more in particular how the estimated \gls{RTF} vectors are utilized to extract a newly activating source
and suppress already active sources.
After defining several projection operators in~\Cref{sec:projection_operators},
\Cref{sec:conventional_methods} introduces the conventional \gls{BOP} method
for successive \gls{RTF} vector estimation.
In \Cref{sec:proposed_methods} we propose three extensions,
offering a closed-form solution to the \gls{BOP} optimization problem,
introducing orthogonal additional vectors and incorporating noise handling.
\Cref{sec:source_activity_estimation} describes the proposed online source counting method based on spatial coherence.
\Cref{sec:evaluation} compares the performance of the proposed method to the conventional \gls{BOP} method
in terms of \gls{RTF} vector estimation accuracy and beamforming performance.
Finally, \Cref{sec:conclusion} concludes the paper by summarizing the contributions and suggesting directions for future research. 

\section{Signal Model}
\label{sec:signal_model}
We consider an acoustic scenario with $K_{\mathrm{max}}$ spatially stationary sound sources
in a noisy and reverberant environment where the sound sources activate successively.
The sound sources are recorded by an array with $M$ microphones,
with $K_{\mathrm{max}} \leq M$. The \gls{STFT} coefficients of the microphone signals at time frame $t$
and frequency bin $f$ are denoted by
\begin{align}
    \mathbf{y}_{t,f} = \begin{bmatrix} y_{1,t,f} & \cdots & y_{M,t,f} \end{bmatrix}^\mathrm{T} \in \mathbb{C}^{M\times 1},
    \label{eq:sig_vec_def}
\end{align}
where $\left\{\cdot\right\}^{\mathrm{T}}$ denotes the transpose operator.
The frequency index $f$ will be omitted for notational brevity (i.e., $\mathbf{y}_{t} = \mathbf{y}_{t,f}$),
unless explicitly required.
At each time frame $t$, the multi-channel microphone signals $\mathbf{y}_t$ can be written as the sum
of the source components $\mathbf{x}_{k,t}$ ($k\in\{1,\ldots,K_{\mathrm{max}}\}$)
and the noise component $\mathbf{n}_t$, i.e.,
\begin{align}
    \mathbf{y}_t
    = \underbrace{\sum_{k=1}^{K_{\mathrm{max}}}\mathcal{I}_{k,t}\mathbf{x}_{k,t}}_{=\mathbf{x}_{t}}
    + \mathbf{n}_t,
    \label{eq:direct_sig_mod_activation}
\end{align}
where the vectors $\mathbf{x}_{k,t}\in\mathbb{C}^{M\times 1}$ and $\mathbf{n}_t\in\mathbb{C}^{M\times 1}$
are defined similarly as in~\Cref{eq:sig_vec_def} and $\mathcal{I}_{k,t}\in\{0,1\}$ denotes the activity of the $k$-th source,
which is not assumed to be frequency-dependent.
In this paper, we consider successively activating sources as depicted in~\Cref{fig:segments}, i.e.,
\begin{align}
    \mathcal{I}_{k,t} = \begin{cases}
                            0, & \mathrm{if}\; t < t_k \\
                            1, & \mathrm{if}\; t \geq t_k
                        \end{cases},
\end{align}
where $t_k$ denotes the activation time of the $k$-th source.
We assume that simultaneous activation of multiple sources does not occur,
i.e., there is at least one frame between the activation of two sources ($t_{k+1}-t_k > 0$).
Assuming the source components and the noise component are uncorrelated,
the noisy covariance matrix
$\mathbf{R}_{y,t}=\mathbb{E}\{\mathbf{y}_{t}\mathbf{y}_{t}^{\mathrm{H}}\}$,
with $\mathbb{E}\{\cdot\}$ denoting the expectation operator
and $\{\cdot\}^{\mathrm{H}}$ denoting the conjugate transpose operator,
can be written as
\begin{align}
    \mathbf{R}_{y,t} = \mathbf{R}_{x,t} + \mathbf{R}_{n},
    \label{eq:Ry_sig_model}
\end{align}
with $\mathbf{R}_{x,t} = \mathbb{E}\{\mathbf{x}_{t}\mathbf{x}_{t}^{\mathrm{H}}\}$
the noiseless covariance matrix
and $\mathbf{R}_{n} = \mathbb{E}\{\mathbf{n}_{t}\mathbf{n}_{t}^{\mathrm{H}}\}$
the noise covariance matrix,
which is assumed to be time-invariant and full-rank.

Assuming sufficiently large \gls{STFT} frames, the $k$-th source component $\mathbf{x}_{k,t}$
can be modeled as the multiplication of the $k$-th source component in a reference microphone, denoted by $r$,
with the (time-invariant) \gls{RTF} vector $\mathbf{g}_{k}\in\mathbb{C}^{M\times 1}$ of the $k$-th source~\cite{avargel_multiplicative_2007}, i.e.,
\begin{align}
    \mathbf{x}_{k,t} = \mathbf{g}_{k}x_{k,r,t}\quad\forall k\in\{1,\ldots,K_{\mathrm{max}}\}.
    \label{eq:direct_sig_mod_RTF}
\end{align}
It should be noted that the reference entry of all \gls{RTF} vectors equals $1$,
i.e., $\mathbf{e}_{r}^{\mathrm{T}}\mathbf{g}_{k} = 1$,
where the $r$-th entry (corresponding to the reference microphone) of the selection vector $\mathbf{e}_{r}$
is equal to $1$ and all other entries are equal to $0$. We assume that the \gls{RTF} vectors of all sources are linearly independent.
The problem of estimating the \gls{RTF} vectors of all sources can be reformulated
into the problem of successively estimating the \gls{RTF} vector of each source
in its activating time segment $\mathcal{T}_k = \left[t_k,t_{k+1}\right[$.
Considering the $K$-th time segment, the corresponding subproblem can be seen as a scenario with $K\leq K_{\mathrm{max}}$ active sources,
where the $K$-th source denotes the newest activating source, i.e.,
\begin{align}
    \mathbf{y}_t = \underbrace{\mathbf{x}_{K,t}}_{\text{new}}
    + \underbrace{\sum_{k=1}^{K-1}\mathbf{x}_{k,t}}_{\text{old}} + \mathbf{n}_t
    = \mathbf{g}_{K}x_{K,r,t}
    + \underbrace{\sum_{k=1}^{K-1}\mathbf{g}_{k}x_{k,r,t}
    + \mathbf{n}_t}_{=\mathbf{v}_{t}},
    \label{eq:direct_sig_mod}
\end{align}
where $\mathbf{v}_{t}\in\mathbb{C}^{M\times 1}$ denotes the undesired component.
By stacking the \gls{RTF} vectors of the already active sources in the matrix
\begin{align}
    \mathbf{G}_{\bar{K}} =
    \begin{bmatrix}
        \mathbf{g}_{1} & \mathbf{g}_{2} & \cdots & \mathbf{g}_{K-1}
    \end{bmatrix}\in\mathbb{C}^{M\times (K-1)},
    \label{eq:RTF_matrix_old_sources}
\end{align}
    the noiseless covariance matrix can be written as
\begin{align}
    \boxed{
    \mathbf{R}_{x,t}
    = \underbrace{\mathbf{g}_{K}\phi_{K,t}\,\mathbf{g}_{K}^{\mathrm{H}}}_{=\mathbf{R}_{K,t}}
    + 
    \underbrace{\mathbf{G}_{\bar{K}}\bm{\Phi}_{\bar{K},t}\mathbf{G}_{\bar{K}}^{\mathrm{H}}}_{=\mathbf{R}_{\bar{K},t}}
    }
    \label{eq:sig_model}
\end{align}
The matrices $\mathbf{R}_{K,t}$ and $\mathbf{R}_{\bar{K},t}$ denote the $M\times M$-dimensional covariance matrices of the $K$-th activating (new) source
and the $K-1$ already active (old) sources, which have rank $1$ and rank $K-1$, respectively.
The diagonal matrix $\bm{\Phi}_{\bar{K},t}=\mathrm{diag}\{\begin{bmatrix}\phi_{1,t} & \phi_{2,t} & \cdots & \phi_{K-1,t}\end{bmatrix}\}\in\mathbb{R}_{+}^{(K-1)\times (K-1)}$
contains the \glspl{PSD} of the already active sources,
with $\mathrm{diag}\{\cdot\}$ building a diagonal matrix from a vector
and $\phi_{k,t}=\mathbb{E}\{\abs{x_{k,r,t}}^{2}\}$ denoting the (time-varying) \gls{PSD} of the $k$-th source component in the reference microphone.

In practice, the covariance matrix $\mathbf{R}_{y,t}$ can be estimated recursively using an exponential sliding window, i.e.,
\begin{align}
    \widehat{\mathbf{R}}_{y,t} = \alpha \widehat{\mathbf{R}}_{y,t-1} + \left(1-\alpha\right)\mathbf{y}_t\mathbf{y}_t^{\mathrm{H}},
    \label{eq:cov_est}
\end{align}
where $\alpha = e^{-\nicefrac{t_{\mathrm{fs}}}{t_{\alpha}}}$ denotes the forgetting factor,
with $t_{\alpha}$ and $t_{\mathrm{fs}}$ the smoothing time constant and the \gls{STFT} frame shift, respectively.
This paper focuses on online estimating the \gls{RTF} vector $\mathbf{g}_{K}$ of the newly activating source,
assuming that estimates of the noise covariance matrix $\mathbf{R}_{n}$
and the \gls{RTF} vectors $\mathbf{G}_{\bar{K}}$ of the already active sources are available.
The \gls{RTF} vectors $\mathbf{g}_{1}\,\ldots\,\mathbf{g}_{K-1}$ can be successively estimated in the first ($K-1$) time segments,
whereas the noise covariance matrix can be estimated in a noise-only segment.
We will first assume that the source activity pattern $\mathcal{I}_{k,t}$ is known a-priori,
while in~\Cref{sec:source_activity_estimation} we will propose an online source counting method.

\section{LCMV beamformer}
\label{sec:LCMV}
As a possible application of successive \gls{RTF} vector estimation,
we will use the estimated \gls{RTF} vectors in an \gls{LCMV} beamformer
to extract the newly activating source and suppress the already active sources
and the noise~\cite{veen_beamforming_1988, hadad_binaural_2016, goessling_blcmv}.
The \gls{LCMV} beamformer minimizes the noise \gls{PSD} subject to linear constraints,
which aim at extracting the $K$-th source component in the reference microphone without distortion
and suppressing the $K-1$ interfering components by a pre-defined amount, i.e.,
\begin{align}
    \mathbf{w}_{K} = \argmin_{\mathbf{w}} \left(\mathbf{w}^{\mathrm{H}}\mathbf{R}_{n}\mathbf{w}\right)
    \quad\mathrm{s.t.}\quad
    \begin{matrix}
        \mathbf{w}^{\mathrm{H}}\mathbf{x}_{K,t} = x_{K,r,t}        \\
        \mathbf{w}^{\mathrm{H}}\mathbf{x}_{k,t} = \delta x_{k,r,t} \\
        \forall k\in \{1,\ldots,K-1\}
    \end{matrix},
    \label{eq:beamforming_opt_problem}
\end{align}
where $\delta\in\left[0,1\right[$ denotes the interference suppression factor (assumed to be equal for all interfering sources).
By reformulating the linear constraints in terms of the \gls{RTF} vectors, i.e., $\mathbf{w}^{\mathrm{H}}\mathbf{g}_{K} = 1$
and $\mathbf{w}^{\mathrm{H}}\mathbf{g}_{k} = \delta\;$ ($\forall k\in \{1,\ldots,K-1\}$),
the well-known solution to the minimization problem in~\Cref{eq:beamforming_opt_problem} is given by
\begin{align}
    \boxed{\mathbf{w}_{K} = \mathbf{R}_{n}^{-1}\mathbf{C}_{K}\left(\mathbf{C}_{K}^{\mathrm{H}}\mathbf{R}_{n}^{-1}\mathbf{C}_{K}\right)^{-1}\bm{\delta}}
    \label{eq:LCMV_beamformer}
\end{align}
where the constraint matrix $\mathbf{C}_{K}\in\mathbb{C}^{M\times K}$ contains the \gls{RTF} vectors of all $K$ active sources,
i.e., $\mathbf{C}_{K} =
\begin{bmatrix}
    \mathbf{g}_K & \mathbf{G}_{\bar{K}}
\end{bmatrix}$,
and $\bm{\delta} = \begin{bmatrix}
        1 & \delta & \cdots & \delta
    \end{bmatrix}^{\mathrm{T}}\in\mathbb{R}_{+}^{K\times 1}$ is the interference suppression vector.
Applying the \gls{LCMV} beamformer to the microphone signals in the $K$-th segment yields the output signal
\begin{align}
    \mathbf{z}_{t} = \mathbf{w}_{K}^{\mathrm{H}}\mathbf{y}_{t}\quad\mathrm{for}\quad t_{K}\leq t < t_{K+1}.
    \label{eq:LCMV_beamforming}
\end{align}
As can be seen in~\Cref{eq:LCMV_beamformer},
the \gls{LCMV} beamformer requires an estimate of the noise covariance matrix $\mathbf{R}_{n}$,
and estimates of the \gls{RTF} vectors of all sources.
Several methods to successively estimate the \gls{RTF} vectors $\mathbf{g}_{k}$ in each segment
will be presented in~\Cref{sec:conventional_methods,sec:proposed_methods}.

\section{Projection Operators}
\label{sec:projection_operators}
This section discusses several projection operators~\cite{banerjee2014linear}
which are fundamental for the \gls{RTF} vector estimation methods in the next sections.
Projection operators, represented by $M\times M$-dimensional projection matrices $\mathbf{P}$ in this context,
project $M$-dimensional vectors onto specific subspaces (see 2D examples in~\Cref{fig:proj_op_scheme}).
All projection matrices are idempotent matrices, i.e., $\mathbf{P}^{2} = \mathbf{P}$.
Hermitian projection matrices, for which $\mathbf{P}^{\mathrm{H}} = \mathbf{P}$, are called orthogonal projection operators,
while all other projection matrices are called oblique projection operators.

\subsubsection{Standard Orthogonal Projection}

\begin{figure}[t!]
    \centering
    \includegraphics[width=0.9\columnwidth]{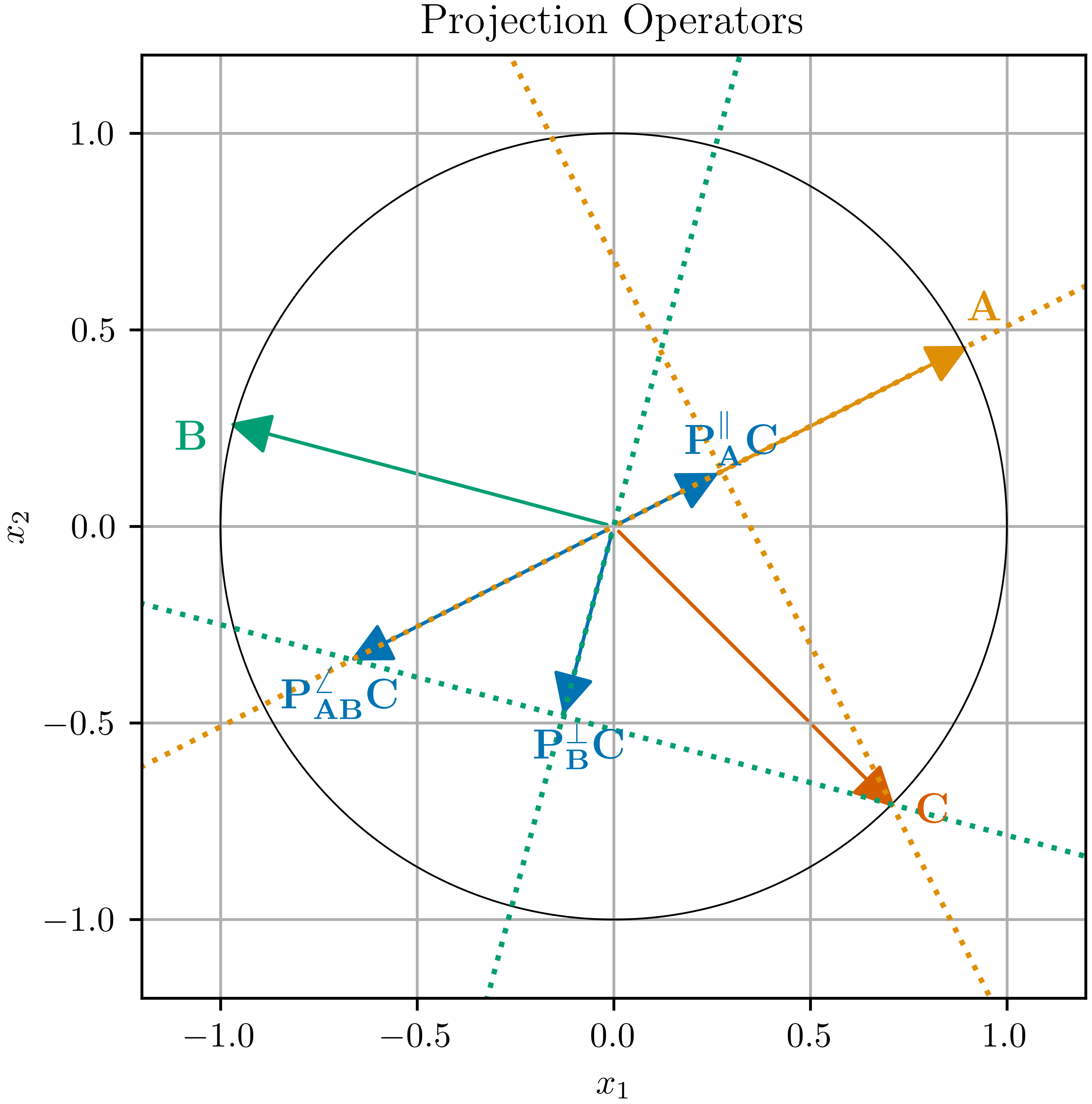}
    \caption{Geometrical interpretation of orthogonal, complement orthogonal and oblique projection in the Euclidean space $\mathbb{R}^{2}$
        based on subspaces spanned by the vectors $\mathbf{A}\in\mathbb{R}^{2\times 1}$ and $\mathbf{B}\in\mathbb{R}^{2\times 1}$, respectively, applied to the vector $\mathbf{C}\in\mathbb{R}^{2\times 1}$.}
    \label{fig:proj_op_scheme}
\end{figure}

The standard orthogonal projection operator $\mathbf{P}^{\parallel}_{\mathbf{A}}$,
projecting vectors onto the column space of the matrix $\mathbf{A}\in\mathbb{C}^{M\times N_{A}}$ ($M\geq N_{A}$, $\text{rank}\left\{\mathbf{A}\right\} = N_{A}$), is defined as:
\begin{align}
    \mathbf{P}^{\parallel}_{\mathbf{A}} = \mathbf{A}\mathbf{A}^{+} = \mathbf{A}\left(\mathbf{A}^{\mathrm{H}}\mathbf{A}\right)^{-1}\mathbf{A}^{\mathrm{H}},
    \label{eq:parallel_projection}
\end{align}
where $\{\cdot\}^{+}$ denotes the Moore-Penrose pseudo-inverse.
Using~\Cref{eq:parallel_projection}, it follows that
\begin{align}
    \mathbf{P}^{\parallel}_{\mathbf{A}}\mathbf{A} = \mathbf{A},
    \quad
    \quad\mathbf{P}^{\parallel}_{\mathbf{A}}\mathbf{A}_{\perp}
    = \mathbf{0}_{M\times (M-N_{A})},
    \label{eq:parallel_projection_properties}
\end{align}
where $\mathbf{0}_{M\times N}$ denotes the $M\times N$-dimensional zero matrix
and $\mathbf{A}_{\perp}\in\mathbb{C}^{M\times (M-N_{A})}$ denotes the orthogonal complement of the matrix $\mathbf{A}$,
which is defined by $\mathbf{A}_{\perp}^{\mathrm{H}} \mathbf{A} = \mathbf{0}_{(M-N_{A}) \times N_{A}}$ and $\text{rank}\left\{
    \begin{bmatrix}
        \mathbf{A} & \mathbf{A}_{\perp}
    \end{bmatrix}
    \right\} = M$.
Since $\mathbf{P}^{\parallel}_{\mathbf{A}}$ is a Hermitian idempotent matrix,
$\mathbf{P}^{\parallel\mathrm{H}}_{\mathbf{A}} \mathbf{P}^{\parallel}_{\mathbf{A}} = \mathbf{P}^{\parallel}_{\mathbf{A}}$.
In the 2D example in~\Cref{fig:proj_op_scheme}, the orthogonal projection operator $\mathbf{P}^{\parallel}_{\mathbf{A}}$ projects the vector $\mathbf{C}$
onto a line through the origin parallel to the vector $\mathbf{A}$, along a line orthogonal to the vector $\mathbf{A}$.

\subsubsection{Complement Orthogonal Projection}
The complement orthogonal projection operator $\mathbf{P}^{\perp}_{\mathbf{A}}$,
projecting vectors onto the subspace orthogonal to the column space of the matrix $\mathbf{A}$, is defined as
\begin{align}
    \mathbf{P}^{\perp}_{\mathbf{A}} 
    = \mathbf{I}_{M} -\mathbf{P}^{\parallel}_{\mathbf{A}},
    \label{eq:orthogonal_projection}
\end{align}
with $\mathbf{I}_{M}$ denoting the $M\times M$-dimensional identity matrix.
From~\Cref{eq:orthogonal_projection}, it follows that
\begin{align}
    \mathbf{P}^{\perp}_{\mathbf{A}}\mathbf{A} = \mathbf{0}_{M\times N_{A}},
    \quad
    \mathbf{P}^{\perp}_{\mathbf{A}}\mathbf{A}_{\perp} = \mathbf{A}_{\perp},
    \\
    \mathrm{rank}\left\{\mathbf{P}^{\perp}_{\mathbf{A}}\right\}
    = M-\mathrm{rank}\left\{\mathbf{A}\right\} = M-N_{A},
    \label{eq:orthogonal_projection_properties}
\end{align}
i.e., every vector in the column space of $\mathbf{A}$ is projected to zero.
Since $\mathbf{P}^{\perp}_{\mathbf{A}}$ is a Hermitian idempotent matrix,
$\mathbf{P}^{\perp\mathrm{H}}_{\mathbf{A}} \mathbf{P}^{\perp}_{\mathbf{A}}
    = \mathbf{P}^{\perp}_{\mathbf{A}}$.
In the 2D example in~\Cref{fig:proj_op_scheme}, the complement orthogonal projection operator $\mathbf{P}^{\perp}_{\mathbf{B}}$ projects the vector $\mathbf{C}$
onto a line through the origin orthogonal to the vector $\mathbf{B}$, along a line parallel to the vector $\mathbf{B}$.

\subsubsection{Oblique Projection}
The oblique projection operator $\mathbf{P}^{\angle}_{\mathbf{AB}}$, projecting vectors onto the column space of the matrix $\mathbf{A}$
while simultaneously projecting vectors in the column space of the matrix $\mathbf{B}\in\mathbb{C}^{M\times N_B}$ ($M\geq N_B$, $\text{rank}\left\{\mathbf{B}\right\} = N_B$)
to zero, is defined as
\begin{align}
    \mathbf{P}^{\angle}_{\mathbf{AB}} =\mathbf{A}\left(\mathbf{A}^{\mathrm{H}}\mathbf{P}^{\perp}_{\mathbf{B}}\mathbf{A}\right)^{-1}\mathbf{A}^{\mathrm{H}}\mathbf{P}^{\perp}_{\mathbf{B}}.
    \label{eq:oblique_projection}
\end{align}
From~\Cref{eq:oblique_projection}, it follows that
\begin{align}
      \mathbf{P}^{\angle}_{\mathbf{AB}}\mathbf{A} 
      = \mathbf{A},
      \quad
      \mathbf{P}^{\angle}_{\mathbf{AB}}\mathbf{B}
      = \mathbf{0}_{M\times N_{B}},
    \label{eq:oblique_projection_properties}
    \\
    \mathbf{P}^{\angle}_{\mathbf{AB}}
    \begin{bmatrix}
        \mathbf{A} & \mathbf{B}
    \end{bmatrix}_{\perp}
    = \mathbf{0}_{M\times (M-N_{A}-N_{B})},
    \label{eq:oblique_projection_properties2}
\end{align}
Since $\mathbf{P}^{\angle}_{\mathbf{AB}} $ is an idempotent matrix,
$\mathbf{P}^{\angle}_{\mathbf{AB}} \mathbf{P}^{\angle}_{\mathbf{AB}} = \mathbf{P}^{\angle}_{\mathbf{AB}}$.
In the 2D example in~\Cref{fig:proj_op_scheme}, the oblique  projection operator $\mathbf{P}^{\angle}_{\mathbf{AB}}$ projects the vector $\mathbf{C}$
onto a line through the origin parallel to the vector $\mathbf{A}$, along a line parallel to the vector $\mathbf{B}$.
Using~\Cref{eq:parallel_projection,eq:orthogonal_projection,eq:oblique_projection},
it can be easily seen that
\begin{align}
    \mathbf{P}^{\angle}_{\mathbf{AB}} = \mathbf{P}^{\parallel}_{\mathbf{A}} \quad \text{if} \quad \mathbf{B} \perp \mathbf{A}.
    \label{eq:oblique_projection_BperpA}
\end{align}
In contrast to the orthogonal projection,
the magnitude of a vector projected by the oblique projection operator
can potentially be larger than the original vector's magnitude
when the angle between the column spaces of $\mathbf{A}$ and $\mathbf{B}$ is small.

\section{Conventional BOP Method}
\label{sec:conventional_methods}
To estimate the \gls{RTF} vector $\mathbf{g}_{K}$ of the source activating in the $K$-th segment,
the \gls{BOP} method in~\cite{cherkassky_successive_2020} aims at blocking this source while keeping the $K-1$ already active sources distortionless.
The \gls{BOP} method applies the oblique projection operator $\mathbf{P}^{\angle}_{\mathbf{G}_{\bar{K}}\bm{\theta}}$ in~\Cref{eq:oblique_projection},
with $\mathbf{G}_{\bar{K}}$ in~\Cref{eq:RTF_matrix_old_sources}
containing the known \gls{RTF} vectors of the already active sources
and $\bm{\theta}$ a vector variable,
to the microphone signals.
The \gls{BOP} cost function is defined as the power of the projected signal, i.e.,
\begin{align}
    J_{\mathbf{G}_{\bar{K}}}\!\!\left(\bm{\theta}\right) \!=\!
    \mathrm{tr}\!\left\{\mathbf{P}^{\angle}_{\mathbf{G}_{\bar{K}}\bm{\theta}}
    \mathbf{R}_{y,t}
    \mathbf{P}^{\angle\mathrm{H}}_{\mathbf{G}_{\bar{K}}\bm{\theta}}\right\},
    \label{eq:proj_sig_model_costfun_def}
\end{align}
where $\mathrm{tr}\{\cdot\}$ denotes the trace.
Using the signal model in~\Cref{eq:sig_model} and the fact that $\mathbf{P}^{\angle}_{\mathbf{G}_{\bar{K}}\bm{\theta}}\mathbf{G}_{\bar{K}} = \mathbf{G}_{\bar{K}}$,
the \gls{BOP} cost function in~\Cref{eq:proj_sig_model_costfun_def} can be written as
\begin{multline}
J_{\mathbf{G}_{\bar{K}}}\!\!\left(\bm{\theta}\right) =
\mathrm{tr}\left\{\mathbf{P}^{\angle}_{\mathbf{G}_{\bar{K}}\bm{\theta}}\mathbf{g}_{K}
\phi_{K,t}\mathbf{g}_{K}^{\mathrm{H}}\mathbf{P}^{\angle\mathrm{H}}_{\mathbf{G}_{\bar{K}}\bm{\theta}}\right\} \\
+
\mathrm{tr}\left\{\mathbf{G}_{\bar{K}}\bm{\Phi}_{\bar{K},t}\mathbf{G}_{\bar{K}}^{\mathrm{H}}\right\}
+ \mathrm{tr}\left\{\mathbf{P}^{\angle}_{\mathbf{G}_{\bar{K}}\bm{\theta}}
\mathbf{R}_{n}
\mathbf{P}^{\angle\mathrm{H}}_{\mathbf{G}_{\bar{K}}\bm{\theta}}\right\},
    \label{eq:proj_sig_model_costfun}
\end{multline}
where the term $\mathrm{tr}\{\mathbf{G}_{\bar{K}}\bm{\Phi}_{\bar{K},t}\mathbf{G}_{\bar{K}}^{\mathrm{H}}\}$ is a positive constant
and
both other terms
depend on the variable $\bm{\theta}$.
The conventional \gls{BOP} method in~\cite{cherkassky_successive_2020} assumes a sufficiently large \gls{SNR},
such that the noise covariance matrix $\mathbf{R}_{n}$ can be neglected in~\Cref{eq:Ry_sig_model}
and the noisy covariance matrix $\mathbf{R}_{y,t}$ reduces to the noiseless covariance matrix $\mathbf{R}_{x,t}$.
Under this assumption, the \gls{BOP} cost function is equal to
\begin{align}
    &J_{\mathbf{G}_{\bar{K}}}\!\!\left(\bm{\theta}\right)
    = \mathrm{tr}\!\left\{\mathbf{P}^{\angle}_{\mathbf{G}_{\bar{K}}\bm{\theta}}
    \mathbf{R}_{x,t}
    \mathbf{P}^{\angle\mathrm{H}}_{\mathbf{G}_{\bar{K}}\bm{\theta}}\right\}
    \label{eq:projected_costfun_noiseless}
    \\
    =& \mathrm{tr}\left\{\mathbf{P}^{\angle}_{\mathbf{G}_{\bar{K}}\bm{\theta}}
    \mathbf{g}_{K}\phi_{K,t}\mathbf{g}_{K}^{\mathrm{H}}
    \mathbf{P}^{\angle\mathrm{H}}_{\mathbf{G}_{\bar{K}}\bm{\theta}}\right\}
    +
    \mathrm{tr}\left\{
        \mathbf{G}_{\bar{K}}\bm{\Phi}_{\bar{K},t}\mathbf{G}_{\bar{K}}^{\mathrm{H}}
        \right\},
    \label{eq:proj_sig_model_costfun_noiseless}
\end{align}
which
is minimized when
$\mathbf{P}^{\angle}_{\mathbf{G}_{\bar{K}}\bm{\theta}}\mathbf{g}_{K} = \mathbf{0}_{M\times 1}$, i.e.,
when $\bm{\theta}$ is equal to $\mathbf{g}_{K}$ (or a scaled version).
Therefore, an estimate of the \gls{RTF} vector $\mathbf{g}_{K}$ of the $K$-th source can be obtained as
\begin{align}
    \boxed{\widehat{\mathbf{g}}_{K}
    = \frac{\tilde{\mathbf{g}}_{K}}{\mathbf{e}_{r}^{\mathrm{T}}\tilde{\mathbf{g}}_{K}}
    \quad\mathrm{with}\quad
    \tilde{\mathbf{g}}_{K} = \argmin_{\bm{\theta}}\left(J_{\mathbf{G}_{\bar{K}}}\!\!\left(\bm{\theta}\right)\right)}
    \label{eq:BOP_min_problem}
\end{align}
Since no closed-form solution for the optimization problem in~\Cref{eq:BOP_min_problem} exists,
it was proposed in~\cite{cherkassky_successive_2020} to use an iterative gradient descent method with an element-wise gradient.

\section{Proposed Extensions}
\label{sec:proposed_methods}
In this section, we propose three extensions to the conventional \gls{BOP} method,
which significantly reduce computational complexity and enhance robustness.
After deriving the vector gradient of the \gls{BOP} cost function
and reviewing the use of additional random vectors in the conventional \gls{BOP} method,
we derive a closed-form solution to the \gls{BOP} cost function in~\Cref{sec:closed_form}.
In~\Cref{sec:orthogonal_addvecs}, we suggest a more appropriate choice of additional vectors instead of random vectors.
In~\Cref{sec:noise_handling}, we generalize the \gls{BOP} method to low \gls{SNR} conditions by integrating noise handling,
motivated by the \acrfull{CS} and \acrfull{CW} methods.
Although the derivations in~\Cref{sec:closed_form,sec:orthogonal_addvecs}
are based on the noiseless covariance matrix (i.e., assuming no noise is present),
the conclusions derived in these sections remain valid when noise is present
if the noise handling proposed in~\Cref{sec:noise_handling} is used.

\subsection{Closed-form solution}
\label{sec:closed_form}
In~\cite{cherkassky_successive_2020} the element-wise gradient
of the \gls{BOP} cost function in~\Cref{eq:projected_costfun_noiseless} was derived.
Instead of the element-wise gradient, here we will consider the vector gradient
since it has a lower computational complexity and is more convenient for later derivations.
The vector gradient of the \gls{BOP} cost function in~\Cref{eq:projected_costfun_noiseless} is equal to
(see detailed derivation in~\Cref{apx:vector_gradient_BOP}):
\begin{multline}
    \nabla_{\bm{\theta}}J_{\mathbf{G}_{\bar{K}}}\!\!\left(\bm{\theta}\right) =
    -\mathbf{P}^{\angle\mathrm{H}}_{\mathbf{G}_{\bar{K}}\bm{\theta}}
    \mathbf{P}^{\angle}_{\mathbf{G}_{\bar{K}}\bm{\theta}}\mathbf{R}_{x,t}
    \mathbf{P}^{\perp}_{\mathbf{G}_{\bar{K}}}\bm{\theta}
    \left(\bm{\theta}^{\mathrm{H}}\mathbf{P}^{\perp}_{\mathbf{G}_{\bar{K}}}\bm{\theta}\right)^{-1}
    \\
    -\mathbf{P}^{\perp}_{\left[\mathbf{G}_{\bar{K}},\bm{\theta}\right]}\mathbf{R}_{x,t}\mathbf{P}^{\angle\mathrm{H}}_{\mathbf{G}_{\bar{K}}\bm{\theta}}\mathbf{G}_{\bar{K}}
    \left(\mathbf{G}_{\bar{K}}^{\mathrm{H}}\mathbf{P}^{\perp}_{\bm{\theta}}\mathbf{G}_{\bar{K}}\right)^{-1}\mathbf{G}_{\bar{K}}^{\mathrm{H}}\bm{\theta}\left(\bm{\theta}^{\mathrm{H}}\bm{\theta}\right)^{-1}.
    \label{eq:BOPgradient}
\end{multline}
Using the signal model in~\Cref{eq:sig_model}
and the fact that for $\bm{\theta} = \mathbf{g}_{K}$ it follows that
$\mathbf{P}^{\angle}_{\mathbf{G}_{\bar{K}}\bm{\theta}}\mathbf{R}_{x,t}
    \mathbf{P}^{\perp}_{\mathbf{G}_{\bar{K}}} = \mathbf{0}_{M\times M}$
and $\mathbf{P}^{\perp}_{\left[\mathbf{G}_{\bar{K}},\bm{\theta}\right]}
\mathbf{R}_{x,t} = \mathbf{0}_{M\times M}$,
it can be easily verified that for $\bm{\theta} = \mathbf{g}_{K}$
the gradient in~\Cref{eq:BOPgradient} is equal to $\mathbf{0}_{M\times 1}$.
However, when $K<M$, the null space of the rank-$K$ matrix
$\begin{bmatrix}
        \mathbf{g}_{K} & \mathbf{G}_{\bar{K}}
    \end{bmatrix}^{\mathrm{H}}
    \in\mathbb{C}^{K\times M}
$
is non-empty
and consists of vectors $\tilde{\bm{\theta}}$ for which
$\mathbf{g}_{K}^{\mathrm{H}}\tilde{\bm{\theta}} = 0$
and $\mathbf{G}_{\bar{K}}^{\mathrm{H}}\tilde{\bm{\theta}} = \mathbf{0}_{(K-1)\times 1}$.
Using~\Cref{eq:oblique_projection_BperpA},
for these vectors $\tilde{\bm{\theta}}$
the oblique projection operator $\mathbf{P}^{\angle}_{\mathbf{G}_{\bar{K}}\bm{\theta}}$
simplifies to the orthogonal projection operator
$\mathbf{P}^{\parallel}_{\mathbf{G}_{\bar{K}}}$,
such that the \gls{BOP} cost function
in~\Cref{eq:proj_sig_model_costfun_noiseless} is equal to
\begin{align}
    J_{\mathbf{G}_{\bar{K}}}\!\!\left(\tilde{\bm{\theta}}\right) =
    \mathrm{tr}\left\{\mathbf{P}^{\parallel}_{\mathbf{G}_{\bar{K}}}
    \mathbf{g}_{K}\phi_{K,t}\mathbf{g}_{K}^{\mathrm{H}}
    \mathbf{P}^{\parallel\mathrm{H}}_{\mathbf{G}_{\bar{K}}}\right\}
    +\mathrm{tr}\left\{
    \mathbf{G}_{\bar{K}}\bm{\Phi}_{\bar{K},t}\mathbf{G}_{\bar{K}}^{\mathrm{H}}
    \right\}.
    \label{eq:BOPcostfun_orthogonal}
\end{align}
Since the cost function in~\Cref{eq:BOPcostfun_orthogonal}
does not depend on $\bm{\theta}$,
for all vectors $\tilde{\bm{\theta}}$ in the null space of
$
\begin{bmatrix}
        \mathbf{g}_{K} & \mathbf{G}_{\bar{K}}
\end{bmatrix}^{\mathrm{H}}
$
the gradient is equal to zero,
revealing a (potentially multidimensional) plateau of the \gls{BOP} cost function
and posing a challenge for gradient descent methods.
To overcome this problem, it was proposed in~\cite{cherkassky_successive_2020}
to add $M-K$ random vectors $\mathbf{G}_{a} = \begin{bmatrix}
        \mathbf{g}_{K+1} & \cdots & \mathbf{g}_{M}
    \end{bmatrix}$ to the \gls{RTF} vectors of the already active sources,
        i.e., $\tilde{\mathbf{G}}_{\bar{K}} = \begin{bmatrix}
        \mathbf{G}_{\bar{K}} & \mathbf{G}_{a}
    \end{bmatrix}$ with $\text{rank}\{\tilde{\mathbf{G}}_{\bar{K}}\} = M-1$,
        so that the matrix $\begin{bmatrix}
        \mathbf{g}_{K} & \tilde{\mathbf{G}}_{\bar{K}}
    \end{bmatrix}$ is full rank and its null space is empty
    (assuming no random vector being in the column space of $\begin{bmatrix}
        \mathbf{g}_{K} & \mathbf{G}_{\bar{K}}
    \end{bmatrix}$).
    Note that introducing these additional random vectors changes
    the \gls{BOP} cost function in~\Cref{eq:projected_costfun_noiseless} 
    and the gradient in~\Cref{eq:BOPgradient} to
    \begin{align}
    J_{\tilde{\mathbf{G}}_{\bar{K}}}\!\!\left(\bm{\theta}\right)\!
    =&
    J_{[
        \mathbf{G}_{\bar{K}}, \mathbf{G}_{a}
    ]}\!\!\left(\bm{\theta}\right)
    =
    \mathrm{tr}\left\{\mathbf{P}^{\angle}_{\tilde{\mathbf{G}}_{\bar{K}}\bm{\theta}}
    \mathbf{R}_{x,t}
    \mathbf{P}^{\angle\mathrm{H}}_{\tilde{\mathbf{G}}_{\bar{K}}\bm{\theta}}\right\},
    \label{eq:proj_sig_model_costfun_addvecs}
    \\
    \nabla_{\bm{\theta}}J_{\tilde{\mathbf{G}}_{\bar{K}}}\!\!\left(\bm{\theta}\right)\!
    =& \!-\!\mathbf{P}^{\angle\mathrm{H}}_{\tilde{\mathbf{G}}_{\bar{K}}\bm{\theta}}\mathbf{P}^{\angle}_{\tilde{\mathbf{G}}_{\bar{K}}\bm{\theta}}\mathbf{R}_{x,t}
    \mathbf{P}^{\perp}_{\tilde{\mathbf{G}}_{\bar{K}}}\!\bm{\theta}\!
    \left(\!\bm{\theta}^{\mathrm{H}}\mathbf{P}^{\perp}_{\tilde{\mathbf{G}}_{\bar{K}}}\!\bm{\theta}\!\right)^{\!-1},
    \label{eq:gradient_with_addvecs}
\end{align}
where the second term of the gradient vanishes since
$\text{rank}\{\begin{bmatrix}
    \tilde{\mathbf{G}}_{\bar{K}} & \bm{\theta}
\end{bmatrix}\} = M$
(except for $\bm{\theta}$ in the column space of $\tilde{\mathbf{G}}_{\bar{K}}$, which is highly improbable),
so that $\mathbf{P}^{\perp}_{\left[\tilde{\mathbf{G}}_{\bar{K}},\bm{\theta}\right]} = \mathbf{0}_{M\times M}$.
Using the signal model in~\Cref{eq:sig_model}
and the fact that for $\bm{\theta} = \mathbf{g}_{K}$
it follows that
$
\mathbf{P}^{\angle}_{\tilde{\mathbf{G}}_{\bar{K}}\bm{\theta}}
\mathbf{R}_{x,t}
\mathbf{P}^{\perp}_{\tilde{\mathbf{G}}_{\bar{K}}}
= \mathbf{0}_{M\times M}
$,
it can be shown that for $\bm{\theta} = \mathbf{g}_{K}$
the gradient in~\Cref{eq:gradient_with_addvecs} is equal to $\mathbf{0}_{M\times 1}$,
so that the cost function with the additional vectors $\mathbf{G}_{a}$ in~\Cref{eq:proj_sig_model_costfun_addvecs}
has the same global minimum as the original cost function in~\Cref{eq:projected_costfun_noiseless}.

Using the vector gradient in~\Cref{eq:gradient_with_addvecs},
we now derive a closed-form solution
minimizing the \gls{BOP} cost function with additional vectors
in~\Cref{eq:proj_sig_model_costfun_addvecs}.
Let us consider the eigenvalue decomposition of the matrix
$\mathbf{R}_{x,t}\mathbf{P}^{\perp}_{\tilde{\mathbf{G}}_{\bar{K}}}\in\mathbb{C}^{M\times M}$, i.e.,
\begin{align}
    \mathbf{R}_{x,t}\mathbf{P}^{\perp}_{\tilde{\mathbf{G}}_{\bar{K}}}\bar{\bm{\theta}}_{m}
    = \lambda_{m}\bar{\bm{\theta}}_{m} \quad\forall m \in \{1,\ldots,M\},
    \label{eq:eigenvalue_decomp}
\end{align}
where $\bar{\bm{\theta}}_{m}$ and $\lambda_{m}$ denote the eigenvectors and eigenvalues, respectively.
Since the rank of $\tilde{\mathbf{G}}_{\bar{K}}$ is equal to $M-1$,
it follows from~\Cref{eq:orthogonal_projection_properties}
that the rank of $\mathbf{P}^{\perp}_{\tilde{\mathbf{G}}_{\bar{K}}}$
and the rank of $\mathbf{R}_{x,t}\mathbf{P}^{\perp}_{\tilde{\mathbf{G}}_{\bar{K}}}$ are equal to $1$.
Hence, all eigenvalues except for the principal eigenvalue $\lambda_1$ are equal to zero,
i.e.,
\begin{align}
    \mathbf{R}_{x,t}\mathbf{P}^{\perp}_{\tilde{\mathbf{G}}_{\bar{K}}}\bar{\bm{\theta}}_{1} = & \lambda_{1}\bar{\bm{\theta}}_{1},                           
    \label{eq:eigenvalue_decomp2}
    \\
    \mathbf{R}_{x,t}\mathbf{P}^{\perp}_{\tilde{\mathbf{G}}_{\bar{K}}}\bar{\bm{\theta}}_{m} = & \mathbf{0}_{M\times 1} \quad\forall m \in \{2,\ldots,M\},
\end{align}
where $\bar{\bm{\theta}}_{1}$ denotes the principal eigenvector.
Using~\Cref{eq:eigenvalue_decomp2},
it can be seen that the principal eigenvector sets the gradient in~\Cref{eq:gradient_with_addvecs} to zero, i.e.,
\begin{equation}
    \begin{aligned}
        \nabla_{\bm{\theta}}J_{\tilde{\mathbf{G}}_{\bar{K}}}\!\!\left(\bar{\bm{\theta}}_{1}\right)
        = & -\mathbf{P}^{\angle\mathrm{H}}_{\tilde{\mathbf{G}}_{\bar{K}}\bar{\bm{\theta}}_{1}}\mathbf{P}^{\angle}_{\tilde{\mathbf{G}}_{\bar{K}}\bar{\bm{\theta}}_{1}}
            \mathbf{R}_{x,t}\mathbf{P}^{\perp}_{\tilde{\mathbf{G}}_{\bar{K}}}\bar{\bm{\theta}}_{1}
        \left(\bar{\bm{\theta}}_{1}^{\mathrm{H}}\mathbf{P}^{\perp}_{\tilde{\mathbf{G}}_{\bar{K}}}\bar{\bm{\theta}}_{1}\right)^{-1}                                    \\
        \!= & -\!\mathbf{P}^{\angle\mathrm{H}}_{\tilde{\mathbf{G}}_{\bar{K}}\bar{\bm{\theta}}_{1}}
        \!\underbrace{\mathbf{P}^{\angle}_{\tilde{\mathbf{G}}_{\bar{K}}\bar{\bm{\theta}}_{1}}\bar{\bm{\theta}}_{1}}_{=\mathbf{0}_{M\times 1}}\!\lambda_{1}
        \!\left(\!\bar{\bm{\theta}}_{1}^{\mathrm{H}}\mathbf{P}^{\perp}_{\tilde{\mathbf{G}}_{\bar{K}}}\bar{\bm{\theta}}_{1}\!\right)^{-1}\!\!.
    \end{aligned}
\end{equation}
Since using the rank-nullity theorem the dimension of the null spaces
of $\mathbf{R}_{x,t}\mathbf{P}^{\perp}_{\tilde{\mathbf{G}}_{\bar{K}}}$
and $\mathbf{P}^{\perp}_{\tilde{\mathbf{G}}_{\bar{K}}}$ are equal to $M-1$,
their null spaces are the same,
so that for all other eigenvectors
$\mathbf{P}^{\perp}_{\tilde{\mathbf{G}}_{\bar{K}}}\bar{\bm{\theta}}_{m} = \mathbf{0}_{M\times 1}$ $(\forall m \in \{2,\ldots,M\})$.
Hence, these eigenvectors can not be solutions
as this would cause a division by zero in the term
$({\bm{\theta}}^{\mathrm{H}}\mathbf{P}^{\perp}_{\tilde{\mathbf{G}}_{\bar{K}}}{\bm{\theta}})^{-1}$
of the gradient in~\Cref{eq:gradient_with_addvecs}.
The closed-form solution minimizing the \gls{BOP} cost function $J_{\tilde{\mathbf{G}}_{\bar{K}}}\!\!(\bm{\theta})$
in~\Cref{eq:proj_sig_model_costfun_addvecs} is hence given by
\begin{align}
    \boxed{\widehat{\mathbf{g}}_{K}
    = \frac{\tilde{\mathbf{g}}_{K}}{\mathbf{e}_{r}^{\mathrm{T}}\tilde{\mathbf{g}}_{K}}
    \quad\mathrm{with}\quad
    \tilde{\mathbf{g}}_{K} = \mathcal{V}_{\mathrm{max}}\left\{\mathbf{R}_{x,t}\mathbf{P}^{\perp}_{\tilde{\mathbf{G}}_{\bar{K}}}\right\}}
    \label{eq:BOP_closed_form}
\end{align}
with $\mathcal{V}_{\mathrm{max}}\{\cdot\}$ denoting the principal eigenvector.

\subsection{Orthogonal Additional Vectors}
\label{sec:orthogonal_addvecs}
As shown in~\Cref{sec:closed_form},
when the signal model in~\Cref{eq:sig_model} holds,
the choice of the $M-K$ additional vectors $\mathbf{G}_{a}$ has no influence
on the global minimum of the \gls{BOP} cost function.
However, since in practice the signal model in~\Cref{eq:sig_model}
and its corresponding assumptions, e.g.,
the rank-1 approximation in~\Cref{eq:direct_sig_mod_RTF} for each source,
may not perfectly hold,
it should be realized that the solution in~\Cref{eq:BOP_closed_form}
depends on the choice of additional vectors $\mathbf{G}_{a}$.
As mentioned in~\Cref{sec:projection_operators},
the oblique projection operator $\mathbf{P}^{\angle}_{\mathbf{A}\mathbf{B}}$
may amplify a vector when the angle between the column spaces
of $\mathbf{A}$ and $\mathbf{B}$ is small.
This means that applying the oblique projection operator
$\mathbf{P}^{\angle}_{\tilde{\mathbf{G}}_{\bar{K}}\bm{\theta}}$
in~\Cref{eq:gradient_with_addvecs} for $\bm{\theta} = \mathbf{g}_{K}$
could potentially lead to amplification of error signal components
caused by model mismatch when the angle between the column spaces
of $\tilde{\mathbf{G}}_{\bar{K}} = 
\begin{bmatrix}
    \mathbf{G}_{\bar{K}} & \mathbf{G_{a}}
\end{bmatrix}$
and $\mathbf{g}_{K}$ is small.
To reduce such amplification, a good choice of additional vectors $\mathbf{G}_{a}$
would hence be vectors which are orthogonal to $\mathbf{g}_{K}$.
Since obviously $\mathbf{g}_{K}$ is not known, 
we propose to compute the $M-K$ additional vectors
as the minor subspace of the covariance matrix $\mathbf{R}_{x,t}$, i.e.,
\begin{align}
    \mathbf{G}_{a} = \mathcal{V}_{\mathrm{min}}^{M-K}\left\{\mathbf{R}_{x,t}\right\},
\end{align}
with $\mathcal{V}_{\mathrm{min}}^{N}\{\cdot\}$ denoting the minor subspace of dimension $N$ spanned by the eigenvectors corresponding to the $N$ smallest eigenvalues.
We call this method \gls{BOPO}.

\subsection{Noise Handling}
\label{sec:noise_handling}
Similarly as in~\cite{cherkassky_successive_2020},
in~\Cref{sec:closed_form,sec:orthogonal_addvecs} it has been assumed
that no noise is present, i.e., all expressions have been derived
using the noiseless covariance matrix $\mathbf{R}_{x,t}$.
Aiming at generalizing the \gls{BOP} method to noisy scenarios,
in this section we propose two noise handling techniques,
which are motivated by the \gls{CS} and \gls{CW} methods
for single-source \gls{RTF} vector
estimation~\cite{warsitz2007blind, markovich_multichannel_2009, serizel_low-rank_2014, varzandeh2017iterative, markovich-golan_performance_2018}.
Both techniques assume an estimate
of the noise covariance matrix $\mathbf{R}_{n}$ to be available
(e.g., estimated from a noise-only segment),
which is used to construct a noiseless covariance matrix
similar to $\mathbf{R}_{x,t}$.
The conventional \gls{BOP} method from~\cite{cherkassky_successive_2020}
assumes that no noise is present,
i.e., $\mathbf{R}_{x,t} = \mathbf{R}_{y,t}$.

\subsubsection{Subtract the noise covariance matrix}
Based on~\Cref{eq:Ry_sig_model}, the noiseless covariance matrix $\mathbf{R}_{x,t}$ can be simply computed
by subtracting the noise covariance matrix $\mathbf{R}_{n}$ from the noisy covariance matrix $\mathbf{R}_{y,t}$, i.e.,
\begin{align}
    \mathbf{R}_{x,t}^{(\mathrm{s})} = \mathbf{R}_{y,t} - \mathbf{R}_{n}.
\end{align}
The corresponding methods are called \gls{BOP-S} and \gls{BOPO-S}, respectively.

\subsubsection{Whiten the noise covariance matrix}
Similarly as for the \gls{CW} method~\cite{markovich-golan_performance_2018},
the noise covariance matrix $\mathbf{R}_{n}$
can also be used to compute the whitened noiseless covariance matrix as
\begin{align}
    \mathbf{R}_{x,t}^{(\mathrm{w})} = \mathbf{R}_{n}^{\nicefrac{-\mathrm{H}}{2}}\mathbf{R}_{y,t}\mathbf{R}_{n}^{\nicefrac{-1}{2}}-\mathbf{I}_{M},
    \label{eq:whitening}
\end{align}
where $\mathbf{R}_{n}^{\nicefrac{1}{2}}$ denotes a matrix square root decomposition
(e.g., Cholesky decomposition) of the noise covariance matrix,
i.e., $\mathbf{R}_{n} = \mathbf{R}_{n}^{\nicefrac{\mathrm{H}}{2}}\mathbf{R}_{n}^{\nicefrac{1}{2}}$.
The whitening operation in~\Cref{eq:whitening} transfers the noisy covariance matrix
into a whitened noiseless domain,
where the same methods as in the original domain can be applied.
However, before adding additional vectors $\mathbf{G}_{a}^{(\mathrm{w})}$,
the \gls{RTF} vectors of the already active sources
need to be transferred into the whitened domain, i.e.,
\begin{align}
    \mathbf{G}_{\bar{K}}^{(\mathrm{w})}
    = \mathbf{R}_{n}^{\nicefrac{-\mathrm{H}}{2}}\mathbf{G}_{\bar{K}},
    \label{eq:whitened_RTFs}
\end{align}
such that $\tilde{\mathbf{G}}_{\bar{K}}^{(\mathrm{w})} = 
\begin{bmatrix}
    \mathbf{G}_{\bar{K}}^{(\mathrm{w})} & \mathbf{G}_{a}^{(\mathrm{w})}
\end{bmatrix}$.
In addition, after using~\Cref{eq:BOP_closed_form} to compute
$\tilde{\mathbf{g}}_{K}^{(\mathrm{w})}
= \mathcal{V}_{\mathrm{max}}\{
    \mathbf{R}_{x,t}^{(\mathrm{w})}\mathbf{P}_{\tilde{\mathbf{G}}_{\bar{K}}^{(\mathrm{w})}}
    \}$
in the whitened domain,
this vector still needs to be transferred into the original domain by de-whitening, i.e.,
\begin{align}
    \tilde{\mathbf{g}}_{K} = \mathbf{R}_{n}^{\nicefrac{\mathrm{H}}{2}}\tilde{\mathbf{g}}_{K}^{(\mathrm{w})}.
    \label{eq:dewhitening}
\end{align}
The reference  entry normalization is then performed in the original domain as shown in~\Cref{eq:BOP_closed_form}.
The corresponding methods are called \gls{BOP-W} and \gls{BOPO-W}, respectively.

\section{Source Activity Estimation}
\label{sec:source_activity_estimation}
Both the conventional \gls{BOP} method and the proposed method require knowledge about the source activity $\mathcal{I}_{k,t}$,
in particular the activation times of newly activating sources.
When a new active source is detected,
the considered successive \gls{RTF} vector estimation methods are triggered
to estimate the \gls{RTF} vector of this new source.
In this section, we propose a spatial-coherence-based online source counting method
to detect new source activations on a frame-by-frame basis.
As mentioned before in this paper, we assume that sources activate successively
and continue to be active.
In practice, the deactivation of a source obviously needs to be handled as well,
which could for example be detected by monitoring a sudden drop in output power
of an \gls{LCMV} beamformer aiming to extract this source.
However, this is beyond the scope of this paper.

The main idea of the proposed online source counting method is
to detect the activation of a single source in spatially white noise
by observing sudden changes in spatial coherence across all frequencies.
The spatial coherence in frequency bin $f$ between the signals captured
by microphone $m$ and $\tilde{m}$ is defined as~\cite{bendat1980engineering}
\begin{align}
    \gamma_{y,t,f}^{m,\tilde{m}}
    = \frac{\mathbb{E}\left\{y_{m,t,f}\,y_{\tilde{m},t,f}^{*}\right\}}
    {\sqrt{\mathbb{E}\left\{\abs{y_{m,t,f}}^2\right\}
    \mathbb{E}\left\{\abs{y_{\tilde{m},t,f}}^2\right\}}},
    \label{eq:spatial_coherence}
\end{align}
where its absolute value ranges between $0$ and $1$.
A localized source exhibits high spatial coherence,
while non-localized (diffuse) sources exhibit low spatial coherence.
Using~\Cref{eq:spatial_coherence}, the $M\times M$-dimensional coherence matrix,
containing the spatial coherence for all microphone pairs,
in frequency bin $f$ and time frame $t$ can be written as
\begin{align}
    \bm{\Gamma}_{y,t,f}
    = \mathbf{D}_{y,t,f}^{-\nicefrac{1}{2}}
    \mathbf{R}_{y,t,f}\mathbf{D}_{y,t,f}^{-\nicefrac{1}{2}},
    \label{eq:coherence_matrix}
\end{align}
where $\mathbf{D}_{y,t,f}$ denotes a diagonal matrix with the same diagonal elements
as the covariance matrix $\mathbf{R}_{y,t,f}$.
As a generalization of the spatial coherence between one microphone pair to multiple microphone pairs,
in~\cite{GMSC_Ramirez_2008} the \gls{GMSC} has been defined as
\begin{align}
    \gamma_{y,t,f} = \frac{\lambda_{\mathrm{max}}\left\{\bm{\Gamma}_{y,t,f}\right\}-1}{M-1},
    \label{eq:GMSC}
\end{align}
where $\lambda_{\mathrm{max}}\left\{\cdot\right\}$ denotes the principal eigenvalue.
It has been shown in~\cite{GMSC_Ramirez_2008} that the \gls{GMSC} $\gamma_{y,t,f}$
ranges between $0$ and $1$.
To obtain a broadband value, a weighted sum over frequencies is computed, i.e.,
\begin{align}
    \gamma_{y,t} = \frac{\sum_{f=1}^{F}w_{f}\gamma_{y,t,f}}{\sum_{f=1}^{F}w_{f}},
    \label{eq:GMSC_weighted}
\end{align}
where $w_{f}$ denotes the weight in frequency bin $f$
and $F$ denotes the number of frequency bins.

Since the activation of the $K$-th source needs to be detected
in the presence of $K-1$ already active sources and background noise,
we propose to use an estimate of the undesired covariance matrix
$\mathbf{R}_{v,t,f} = \mathbb{E}\{\mathbf{v}_{t,f}\mathbf{v}_{t,f}^{\mathrm{H}}\} = \mathbf{R}_{\bar{K},t,f} + \mathbf{R}_{n,f}$
(see signal model in~\Cref{eq:direct_sig_mod})
to whiten the microphone signals at each time frame $t$, i.e.,
\begin{align}
    \mathbf{R}_{w,t,f}
    = \mathbf{R}_{v,t,f}^{\nicefrac{-\mathrm{H}}{2}}
    \mathbf{R}_{y,t,f}\mathbf{R}_{v,t,f}^{\nicefrac{-1}{2}}.
  \label{eq:whitening2n}
\end{align}
As long as the $K$-th source is not active, i.e., in the segment $[t_{K-1},t_{K}[$,
$\mathbf{R}_{y,t,f} = \mathbf{R}_{v,t,f}$ and
the whitened covariance matrix $\mathbf{R}_{w,t,f}$ is equal to $\mathbf{I}_{M}$,
such that the \gls{GMSC} in~\Cref{eq:GMSC} is equal to $0$.
When the $K$-th source activates, the whitened covariance matrix $\mathbf{R}_{w,t,f}$ becomes equal to
$\mathbf{R}_{v,t,f}^{\nicefrac{-\mathrm{H}}{2}}
\mathbf{R}_{K,t,f}\mathbf{R}_{v,t,f}^{\nicefrac{-1}{2}}+\mathbf{I}_{M}$,
with $\mathbf{R}_{v,t,f}^{\nicefrac{-\mathrm{H}}{2}}
\mathbf{R}_{K,t,f}\mathbf{R}_{v,t,f}^{\nicefrac{-1}{2}}$
a rank-$1$ matrix according to the signal model in~\Cref{eq:sig_model}.
This means that the \gls{GMSC} becomes larger than $0$,
with its value depending on the \gls{SNR} of the $K$-th source~\cite{Habets_on_the_spatial_coherence_2012}.
As an estimate of the covariance matrix $\mathbf{R}_{v,t,f}$
of the $K-1$ already active sources and the background noise at time frame $t$,
we propose to use the noisy covariance matrix from $t_{v}$ frames earlier, i.e.,
\begin{align}
    \widehat{\mathbf{R}}_{v,t,f} = \mathbf{R}_{y,t-t_{v},f}\;.
    \label{eq:estimate_Rv}
\end{align}
The number of frames $t_{v}=\nicefrac{t_{\mathrm{sad}}}{t_{\mathrm{fs}}}$
corresponds to the assumed minimal time interval $t_{\mathrm{sad}}$
between the activation of two sources.
\begin{figure}
    \centering
    \includegraphics[width=1\columnwidth]{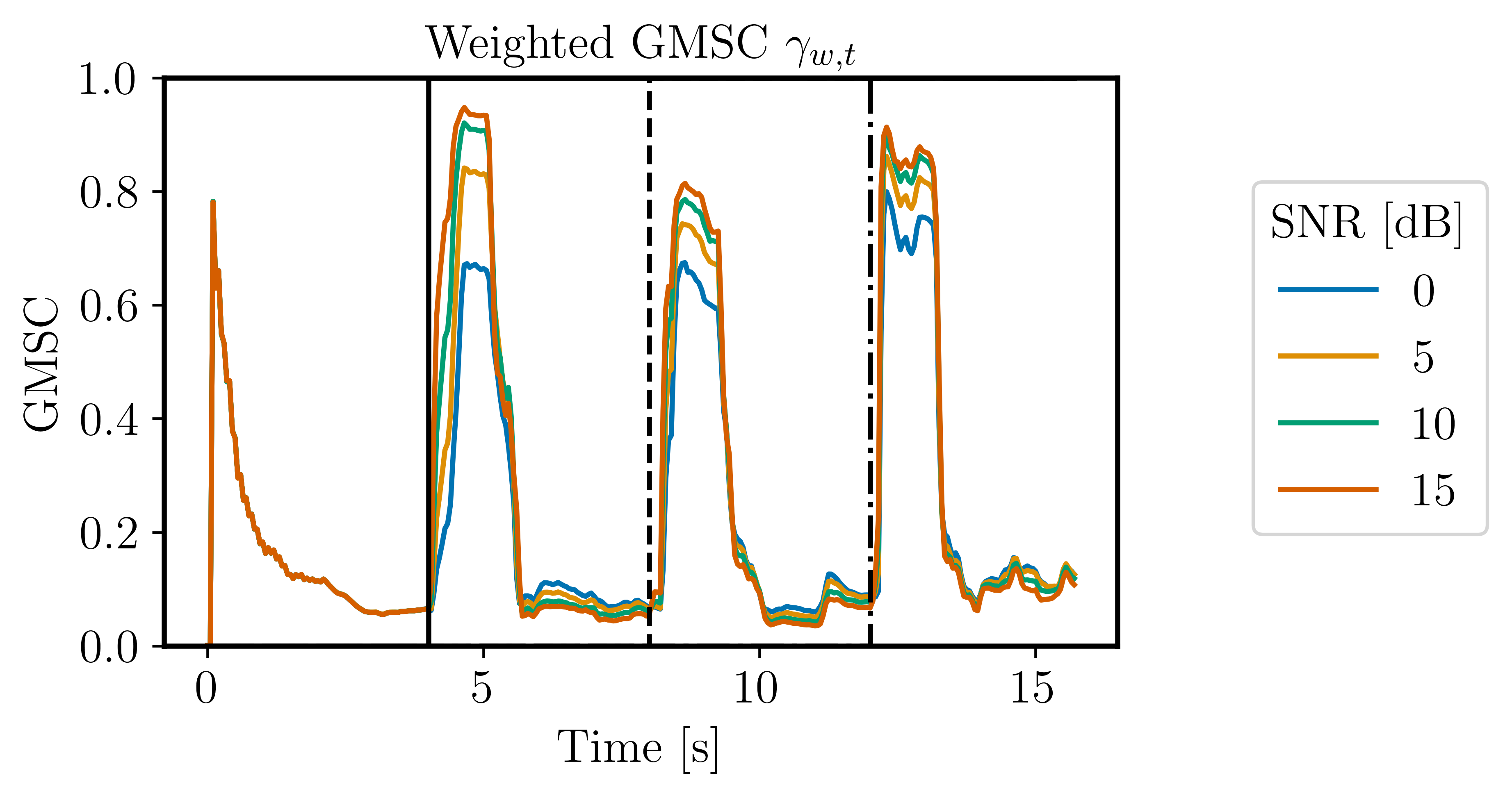}
    \caption{Example of the weighted \gls{GMSC} $\gamma_{w,t}$ over time frames $t$
        for source activations at $\SI{4}{\s}$, $\SI{8}{\s}$ and $\SI{12}{\s}$ and different \glspl{SNR}.}
    \label{fig:GMSC_Measure}
\end{figure}
We propose to compute the weighted \gls{GMSC} $\gamma_{w,t}$
similarly as in~\Cref{eq:GMSC_weighted}
using the whitened covariance matrix $\mathbf{R}_{w,t,f}$ instead of $\mathbf{R}_{y,t,f}$
in~\Cref{eq:coherence_matrix},
and using the power
of the whitened signal as frequency weights, 
i.e., $w_{f} = \mathrm{tr}\left\{\mathbf{R}_{w,t,f}\right\}$.
The weighted \gls{GMSC} is then computed as
\begin{align}
     \gamma_{w,t} = \frac{\sum_{f=1}^{F}\mathrm{tr}\left\{\mathbf{R}_{w,t,f}\right\}
    \frac{
        \lambda_{\mathrm{max}}\left\{
        \mathbf{D}_{w,t,f}^{-\nicefrac{1}{2}}
    \mathbf{R}_{w,t,f}\mathbf{D}_{w,t,f}^{-\nicefrac{1}{2}}
        \right\}-1
        }{M-1}
     }{ 
     \sum_{f=1}^{F}\mathrm{tr}\left\{\mathbf{R}_{w,t,f}\right\}}.
    \label{eq:GMSC_weighted_whitened}
\end{align}
For the scenario depicted in~\Cref{fig:segments},
\Cref{fig:GMSC_Measure} illustrates the weighted \gls{GMSC} $\gamma_{w,t}$
over time frames $t$ for four different \glspl{SNR}
(same \gls{SNR} for each source).
As can be observed, for all \glspl{SNR}
the weighted \gls{GMSC} rises quickly after the activation of a source,
with lower \gls{GMSC} values for lower \glspl{SNR}.
To increase robustness,
the weighted \gls{GMSC} is smoothed
using an exponential window, i.e.,
\begin{align}
    \bar{\gamma}_{w,t} = \beta \bar{\gamma}_{w,t-1} + (\beta-1) \gamma_{w,t},
    \label{eq:smoothed_GMSC}
\end{align}
where $\beta = e^{-\nicefrac{t_{\mathrm{fs}}}{t_{\gamma}}}$ denotes the forgetting factor,
with $t_{\gamma}$ the smoothing time constant.
The activation of a new source is detected
when the smoothed \gls{GMSC} $\bar{\gamma}_{w,t}$ exceeds a pre-defined threshold $\gamma_{\tau}$.

\section{Evaluation}
\label{sec:evaluation}
In this section, we compare the performance
of the proposed successive \gls{RTF} vector estimation method
using the extensions discussed in~\Cref{sec:proposed_methods}
to the conventional \gls{BOP} method.
After introducing the acoustic scenario
with three successively activating sources in~\Cref{sec:acoustic_scenario},
algorithmic implementation details
and computational complexity are discussed in~\Cref{sec:algorithmic_details}.
\Cref{sec:metrics} discusses the considered performance metrics,
namely \gls{RTF} vector estimation accuracy and \gls{SINR} improvement of an \gls{LCMV} beamformer.
The evaluation is divided into two parts:
in~\Cref{sec:results_oracle} oracle knowledge of source activity is assumed,
whereas in~\Cref{sec:results_online_SAD}
the online source counting method proposed in~\Cref{sec:source_activity_estimation} is used
to validate the performance under more realistic conditions.

\subsection{Acoustic Scenario}
\label{sec:acoustic_scenario}
\begin{figure}
    \centering
    \includegraphics[width=1\columnwidth]{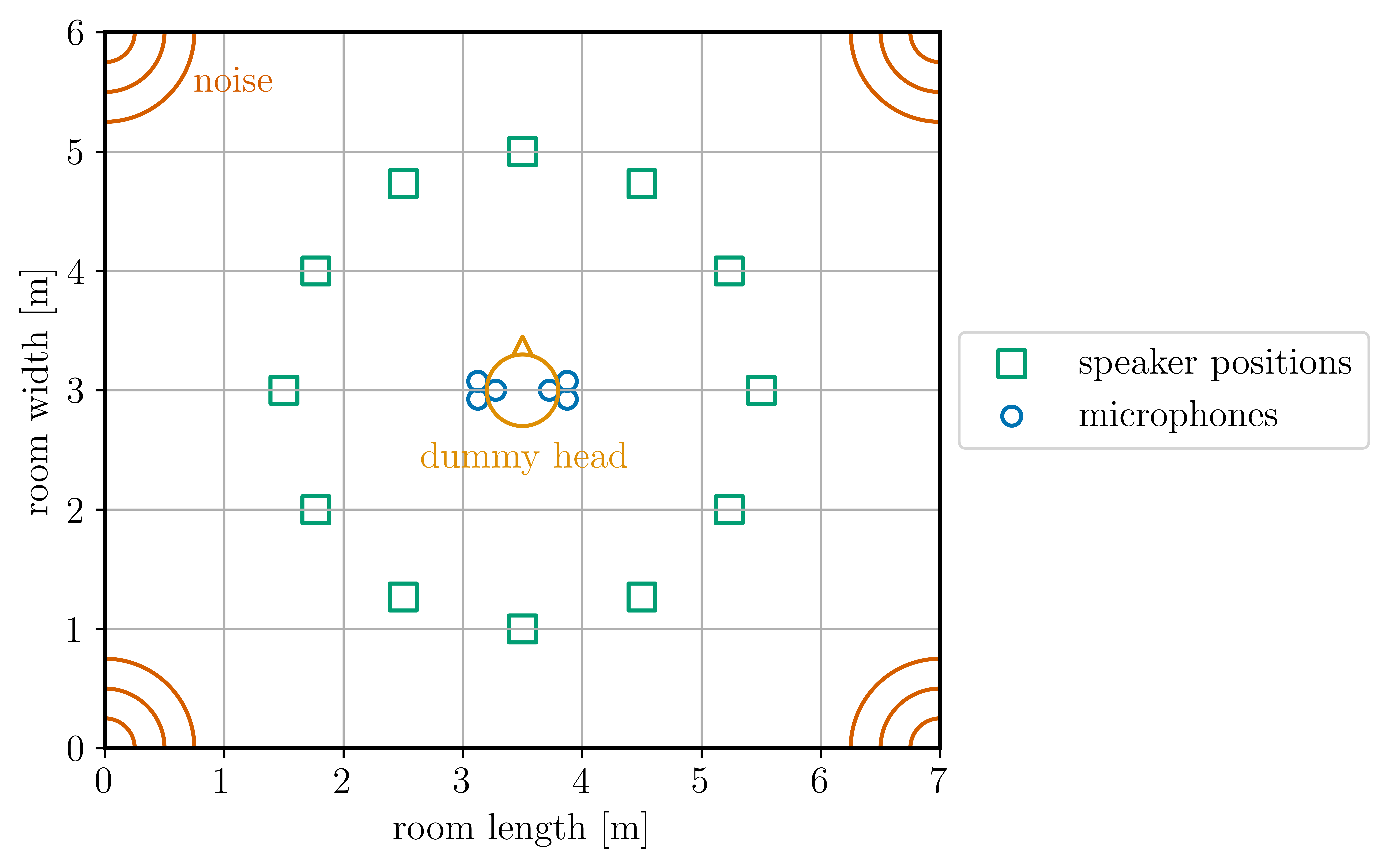}
    \caption{Acoustic setup with 12 speaker positions,
        2 behind-the-ear hearing aids with two microphones each worn by a dummy head,
        one in-ear microphone on each side of the dummy head,
        and background noise from the BRUDEX database~\cite{fejgin2023brudex}.}
    \label{fig:acoustic_scene}
\end{figure}
To generate noisy and reverberant microphone signals,
we used measured \glspl{RIR} and noise signals from the BRUDEX database~\cite{fejgin2023brudex}
at a sampling frequency of $\SI{16}{\kilo\hertz}$.
The acoustic setup is illustrated in~\Cref{fig:acoustic_scene}.
A dummy head was positioned approximately in the center of an acoustic laboratory
measuring $\SI{7}{\m}\times \SI{6}{\m} \times \SI{2.7}{\m}$,
with a reverberation time $T_{60}\approx \SI{310}{\milli\s}$.
The dummy head was equipped with left and right behind-the-ear hearing aids,
each featuring two microphones, and one in-ear microphone on each side,
resulting in $M=6$ microphones.
Each acoustic scenario lasted $\SI{16}{\s}$
and involved $K_{\mathrm{max}} = 3$ speakers of random sex,
activating successively at $\SI{4}{\s}$, $\SI{8}{\s}$ and $\SI{12}{\s}$,
and continuously active background noise (similarly as in~\Cref{fig:segments}).
The source components at the microphones were generated
by convolving clean speech signals from the DNS Challenge~\cite{reddy2020interspeech}
with \glspl{RIR} measured from loudspeakers
placed at 12 different positions around the dummy head in a circle with a radius of $\SI{2}{\m}$
and an angular spacing of $\SI{30}{\degree}$ (see~\Cref{fig:acoustic_scene}).
Quasi-diffuse noise was generated by playing back uncorrelated babble noise or cafeteria noise
using four loudspeakers facing the corners of the laboratory.
Using only the active samples for each source,
the sources were normalized to have equal power,
averaged across all microphones and the entire acoustic scenario.
The noise was then scaled to set the \gls{SNR},
which is defined as the power ratio between a single source component and the noise component.
We considered all possible combinations of source positions and the two mentioned noise types,
leading to 264 evaluated scenarios in the dual-source segment and 2640 evaluated scenarios in the triple-source segment, respectively.

\begin{table}[t!]
    \centering
    \begin{tabular}{lccc}
        \toprule
        \multirow{2}{*}{\textbf{Additional Vectors}} & \multicolumn{3}{c}{\textbf{Noise Handling}}                                 \\ \cline{2-4}
                                                     & no                                          & subtraction   & whitening     \\ \midrule
        random                                       & \gls{BOP}                                   & \gls{BOP-S}*  & \gls{BOP-W}*  \\
        orthogonal                                   & \gls{BOPO}*                                 & \gls{BOPO-S}* & \gls{BOPO-W}* \\
        \bottomrule
    \end{tabular}
    \vspace{0.1cm}
    \caption{Overview of considered \gls{RTF} vector estimation methods,
    resulting from combinations of noise handling and additional vectors.
        Here \gls{BOP} refers to the conventional method~\cite{cherkassky_successive_2020}
        and * denotes the proposed methods.}
    \label{tab:algorithms_overview}
\end{table}

\subsection{Algorithmic Implementation}
\label{sec:algorithmic_details}
The algorithms were implemented using an \gls{STFT} framework
with a frame length of 3200 samples (corresponding to $\SI{200}{\milli\s}$),
a frame shift of 800 samples (corresponding to $t_{\mathrm{fs}}=\SI{50}{\milli\s}$)
and a square-root-Hann window for analysis and synthesis.
The noise covariance matrix is estimated as the sample covariance matrix in the noise-only segment $\mathcal{T}_{n} = [0,t_{1}[$,
i.e., $\widehat{\mathbf{R}}_{n}
= \nicefrac{1}{\abs{\mathcal{T}_{\mathrm{n}}}}\sum_{t\in\mathcal{T}_{\mathrm{n}}}\mathbf{y}_{t}\mathbf{y}_{t}^{\mathrm{H}}$.
The noisy covariance matrix $\widehat{\mathbf{R}}_{y,t}$ is initialized
in the first time frame and then updated recursively
using~\Cref{eq:cov_est} with a smoothing time constant of $t_{\alpha} = \SI{1}{\s}$.
Similarly as in~\Cref{eq:whitening} and~\Cref{eq:dewhitening},
the \gls{RTF} vector of the first source is estimated
as the de-whitened principal eigenvector of the whitened noisy covariance matrix, i.e.,
\begin{align}
    \widehat{\mathbf{g}}_{1,t} =& \frac{\mathbf{R}_{n}^{\nicefrac{\mathrm{H}}{2}}\tilde{\mathbf{g}}_{1,t}}
    {\mathbf{e}_{r}^{\mathrm{T}}\mathbf{R}_{n}^{\nicefrac{\mathrm{H}}{2}}\tilde{\mathbf{g}}_{1,t}}
    \\
    \quad\text{with}\quad
    \tilde{\mathbf{g}}_{1,t}
    =&\mathcal{V}_{\mathrm{max}}
    \left\{\mathbf{R}_{n}^{\nicefrac{-\mathrm{H}}{2}}
    \mathbf{R}_{y,t}\mathbf{R}_{n}^{\nicefrac{-1}{2}} - \mathbf{I}_{M}\right\}.
    \label{eq:RTF_estimation_first_source}
\end{align}
The \gls{RTF} vectors of the second and third sources
$\widehat{\mathbf{g}}_{2,t}$ and $\widehat{\mathbf{g}}_{3,t}$ are estimated
in the dual-source and triple-source segments, respectively.
\Cref{tab:algorithms_overview} provides an overview
of the six considered \gls{RTF} vector estimation methods,
resulting from the combination of three possible noise handling techniques (no, subtraction, whitening)
and two choices of additional vectors (random, orthogonal).
The matrix $\mathbf{G}_{\Bar{K}}$ is built
using the estimated \gls{RTF} vectors of the already active sources
from the last frame of each source's corresponding segment,
i.e., $\mathbf{G}_{2} = \widehat{\mathbf{g}}_{1,(t_{2}-1)}$
and $\mathbf{G}_{3}
= \left[\widehat{\mathbf{g}}_{1,(t_{2}-1)},
\widehat{\mathbf{g}}_{2,(t_{3}-1)}\right]$.
Note that in the triple-source segment, each method uses its own estimate
of the \gls{RTF} vector $\widehat{\mathbf{g}}_{2,(t_{3}-1)}$ of the second source
to estimate the \gls{RTF} vector $\widehat{\mathbf{g}}_{3,t}$ of the third source.

For all \gls{RTF} vector estimation methods (including \gls{BOP}),
we utilized the proposed closed-form solution in~\Cref{eq:BOP_closed_form},
significantly reducing computation time
compared to the iterative gradient descent optimization scheme
used in~\cite{cherkassky_successive_2020}.
In addition, the closed-form solution guarantees
computing the global minimum, whereas gradient descent may converge to a local minimum
and depends on multiple parameters,
such as initialization, learning rate, and re-initialization scheme.
Moreover, the closed-form solution is easily parallelizable across frequencies,
unlike gradient descent.
On an NVIDIA RTX A6000 graphics card,
the proposed closed-form solution achieves an overall computation time speed-up
of approximately $1e^6$.

\subsection{Performance Metrics}
\label{sec:metrics}

The \gls{RTF} vector estimation accuracy is evaluated
using the Hermitian angle between the estimated \gls{RTF} vectors and the ground-truth \gls{RTF} vectors.
The Hermitian angle measures the directional similarity between two vectors
and is invariant under arbitrary complex scaling of either vector.
Similarly as in~\Cref{eq:GMSC_weighted}, 
a weighted sum over all $F$ frequency bins and all time frames $t\in\mathcal{T}_{K}$ is computed
($K\in\{1,2,3\}$), i.e.,
\begin{align}
    \psi_{K} =
    \frac{\sum_{t\in\mathcal{T}_{K}}\sum_{f=1}^{F}w_{K,f}\arccos
    \left\{\frac{\abs{\mathbf{g}_{K,f}^{\mathrm{H}}\widehat{\mathbf{g}}_{K,t,f}}}
    {\norm{\mathbf{g}_{K,f}}\norm{\widehat{\mathbf{g}}_{K,t,f}}}\right\}}
    {\abs{\mathcal{T}_{K}}\sum_{f=1}^{F}w_{K,f}},
    \label{eq:WMHA}
\end{align}
where the frequency weights are set to the average signal power of the $K$-th source,
i.e.,  $w_{K,f} = \mathrm{tr}\{\mathbf{R}_{K,f}\}$ with
$\mathbf{R}_{K,f} = \nicefrac{1}{\abs{\mathcal{T}_K}}
\sum_{t\in\mathcal{T}_K}\mathbf{x}_{K,t,f}\mathbf{x}_{K,t,f}^{\mathrm{H}}$.
This weighting ensures that Hermitian angles in frequency bins with low signal power,
which are less relevant for signal enhancement, are weighted less.
The vectors $\mathbf{g}_{K,f}$ and $\widehat{\mathbf{g}}_{K,t,f}$ denote the ground-truth
and estimated \gls{RTF} vector in frequency bin $f$, respectively.
The ground-truth \gls{RTF} vector of the $K$-th source is computed
as the principal eigenvector of its
sample covariance matrix,
i.e., $\mathbf{g}_{K,f} = \mathcal{V}_{\mathrm{max}}\left\{\mathbf{R}_{K,f}\right\}$.
Note that lower values of the weighted Hermitian angle in~\Cref{eq:WMHA}
indicate better performance.

In addition to \gls{RTF} vector estimation accuracy,
we evaluate performance
in terms of the multi-channel broadband \acrfull{SINR} improvement of an \gls{LCMV} beamformer,
aiming at extracting the newly activating source
and suppressing the already active sources and background noise (see~\Cref{sec:LCMV}).
In the time-domain we define the microphone signals
$\mathbf{y}_{l}\in\mathbb{R}^{M\times 1}$,
the $K$-th source components $\mathbf{x}_{K,l}\in\mathbb{R}^{M\times 1}$
and the $K$-th undesired components
$\mathbf{v}_{\Bar{K},l} = \mathbf{y}_{l}-\mathbf{x}_{K,l}$, with $l$ denoting the time-domain sample index.
The \gls{LCMV} beamformer 
in~\Cref{eq:LCMV_beamformer}
with $\delta=-\SI{20}{\dB}$ is applied 
to the $K$-th source component $\mathbf{x}_{K,t}$
and the $K$-th undesired component $\mathbf{v}_{t}$
in the \gls{STFT} domain,
for each reference channel $r\in\{1,\ldots,M\}$.
The corresponding multi-channel time-domain output signals
$\mathbf{x}_{K,l}^{\mathrm{out}}\in\mathbb{R}^{M\times 1}$
and $\mathbf{v}_{\Bar{K},l}^{\mathrm{out}}\in\mathbb{R}^{M\times 1}$
are computed by applying the inverse \gls{STFT} in an overlap-add procedure.
The multi-channel broadband \gls{SINR} improvement is defined as
\begin{multline}
    \Delta \mathrm{SINR}_K = \\ 
    10\log_{10}\!\left\{\! \frac{\!
    \sum_{l\in\mathcal{L}_{K}} \norm{\mathbf{x}_{K,l}^{\mathrm{out}}}^2\!}
    {\!\sum_{l\in\mathcal{L}_{K}} \norm{\mathbf{v}_{\Bar{K},l}^{\mathrm{out}}}^2\!}\!\right\}\!
    -
    10\log_{10}\!\left\{\! \frac{\!
    \sum_{l\in\mathcal{L}_{K}} \norm{\mathbf{x}_{K,l}}^2\!}
    {\!\sum_{l\in\mathcal{L}_{K}} \norm{\mathbf{v}_{\Bar{K},l}}^2\!}\!\right\}\!,
    \label{eq:SINR_improvement}
    \end{multline}
where $\mathcal{L}_{K}$ denotes the set of samples
within the $K$-th segment where all corresponding $K$ sources are active,
determined using a power-based \gls{VAD} with a threshold of $\SI{-30}{\dB}$.

\subsection{Results using oracle source activity knowledge}
\label{sec:results_oracle}

\begin{figure*}[!t]
    \centering
    \includegraphics[width=1\linewidth]{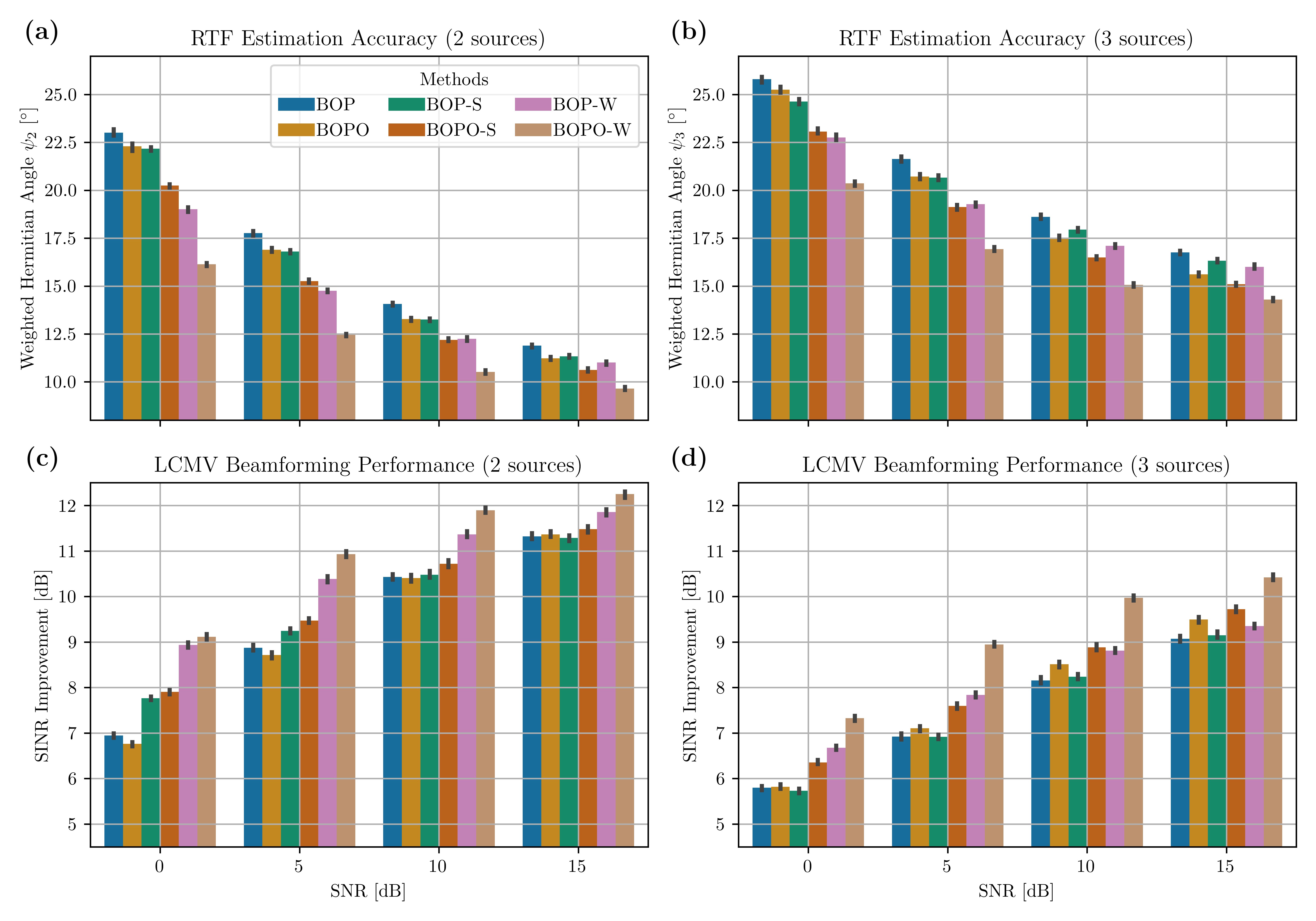}
    \caption{Performance evaluation of the conventional and proposed \gls{RTF} vector estimation methods
    for different \glspl{SNR}     
    using oracle source activity knowledge
        in terms of \gls{RTF} vector estimation accuracy and \gls{SINR} improvement.
        (a) and (c) show the performance for the dual-source segments, 
        while (b) and (d) show the performance for the triple-source segments.}
    \label{fig:results}
\end{figure*}

In this section, we investigate the benefit of the extensions
proposed in~\Cref{sec:orthogonal_addvecs,sec:noise_handling}
compared to the conventional \gls{BOP} method,
assuming oracle source activity knowledge.
For different \glspl{SNR},~\Cref{fig:results} depicts the weighted Hermitian angle
for all considered \gls{RTF} vector estimation methods (see \Cref{tab:algorithms_overview})
and the \gls{SINR} improvement
in dual-source and triple-source segments.
The barplots show the median over all evaluated scenarios
(264 in the dual-source segment and 2640 in the triple-source segment)
with confidence interval ($p=0.05$).
First, it can be observed that in general the performance of all considered methods
improves (i.e., smaller weighted Hermitian angle and larger \gls{SINR} improvement)
with increasing \gls{SNR}, both for the dual-source segments and triple-source segments.
In addition, for all \glspl{SNR} all considered methods perform better
in the dual-source segments compared to the triple-source segments.
Second, we compare the performance between using
orthogonal additional vectors (\gls{BOPO} methods) and random additional vectors (\gls{BOP} methods).
For all \glspl{SNR} and noise handling techniques,
the results for the dual-source segments in~\Cref{fig:results} (a)
and the triple-source segments in~\Cref{fig:results} (b) show
that orthogonal additional vectors
outperform random additional vectors in terms of weighted Hermitian angle,
especially at low \glspl{SNR}.
The advantage of orthogonal over random additional vectors is also observed
in terms of \gls{SINR} improvement, see~\Cref{fig:results} (c) and (d),
except for methods using no noise handling (\gls{BOP} and \gls{BOPO}) in the dual-source segments.
Third, we compare the performance between different noise handling techniques.
Incorporating noise handling for the baseline \gls{BOP} method
with random additional vectors significantly improves performance
(i.e. smaller weighted Hermitian angle and larger \gls{SINR} improvement),
where noise whitening clearly outperforms noise subtraction.
This performance improvement is more pronounced for low \gls{SNR} conditions,
both in the dual-source and triple-source segments.
Also when using orthogonal additional vectors,
noise handling significantly improves performance,
especially for low \gls{SNR} conditions,
where noise whitening outperforms noise subtraction.
In conclusion, the results in \Cref{fig:results} show
that the \gls{BOPO-W} method combining orthogonal additional vectors and noise whitening
provides the best performance of all considered \gls{RTF} vector estimation methods
in terms of \gls{RTF} vector estimation accuracy and \gls{SINR} improvement.
For example, for the challenging $\SI{0}{\dB}$ \gls{SNR} condition in the triple-source segments,
the proposed \gls{BOPO-W} method yields a median improvement of $\SI{5.4}{\degree}$
in terms of weighted Hermitian angle and $\SI{1.5}{\dB}$ in terms of \gls{SINR} improvement
compared to the conventional \gls{BOP} method.

\begin{figure*}[!t]
    \centering
    \includegraphics[width=1\linewidth]{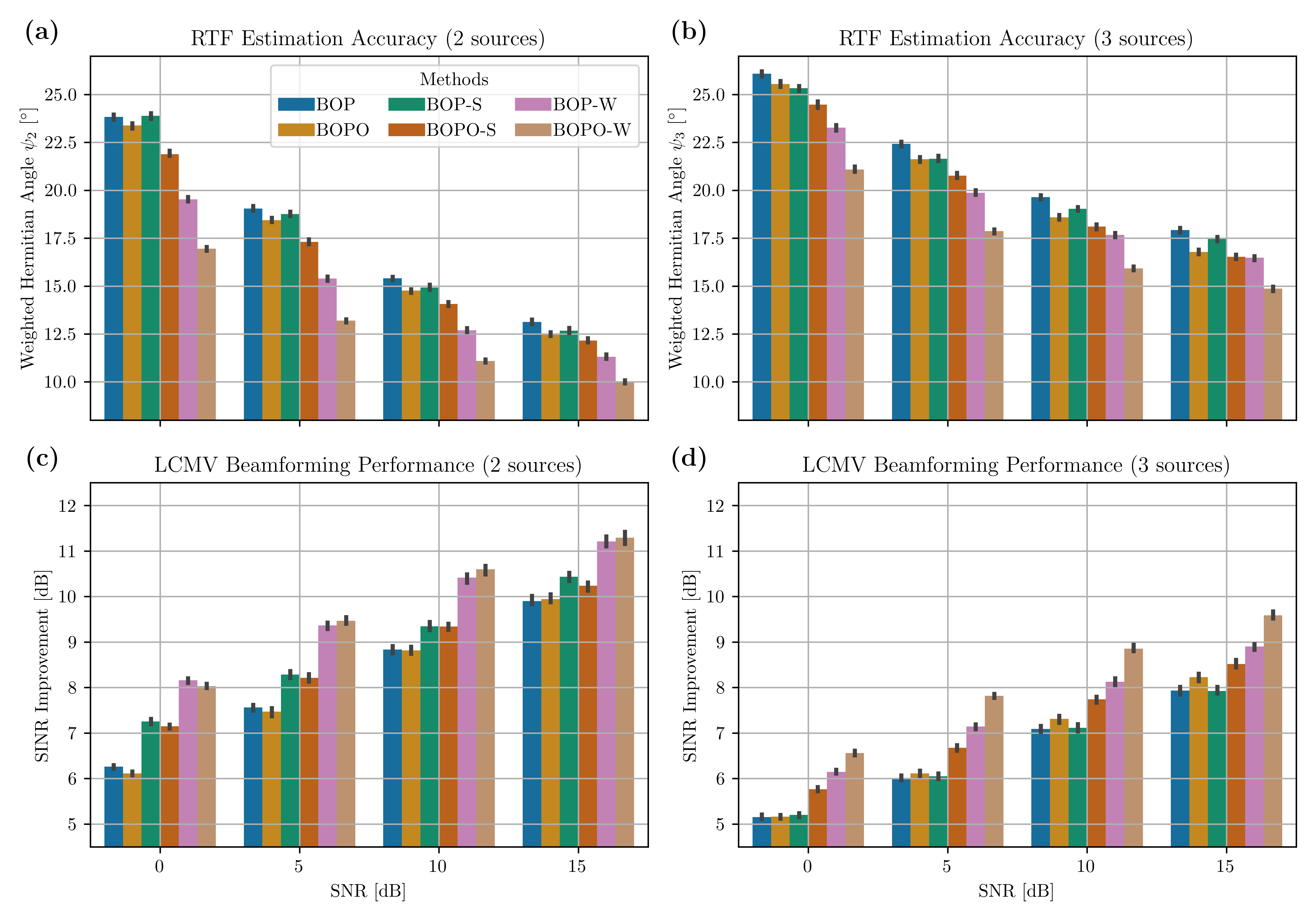}%
    \caption{Performance evaluation of the conventional and proposed \gls{RTF} vector estimation methods
    for different \glspl{SNR}     
    using online source counting method
        in terms of \gls{RTF} vector estimation accuracy and \gls{SINR} improvement.
        (a) and (c) show the performance for the dual-source segments, 
        while (b) and (d) show the performance for the triple-source segments.}
        \label{fig:SC_results}
\end{figure*}

\subsection{Results using Online Source Counting}
\label{sec:results_online_SAD}
Instead of assuming oracle source activity knowledge,
in this section we compare the performance of the proposed methods
and the conventional \gls{BOP} method
using the online source counting method proposed in~\Cref{sec:source_activity_estimation},
which estimates the activation times of the sources on a frame-by-frame basis.
The time constants $t_{\mathrm{sad}}$ and $t_{\gamma}$
in~\Cref{eq:estimate_Rv,eq:smoothed_GMSC} are both set to $\SI{1}{s}$
(same as smoothing constant $t_\alpha$),
assuming that no activation of two sources occurs within a one second interval.
The spatial coherence threshold is set to $\gamma_{\tau}=0.2$.

For different \glspl{SNR}, \Cref{fig:SC_results} depicts the weighted Hermitian angle
and the \gls{SINR} improvement in dual-source and triple-source segments.
First, it can be observed that the performance for all considered methods and conditions
is only slightly lower when using the \gls{GMSC}-based online source counting method
compared to assuming oracle source activity knowledge.
For the weighted Hermitian angle,
the maximal degradation is \SI{1.0}{\degree}
and for the \gls{SINR} improvement, the maximal degradation is \SI{1.5}{\dB}.
These results indicate that the proposed \gls{GMSC}-based online source counting method is able
to provide a good estimate of the source activity pattern.
The overall performance trends 
when using the online source counting method
are very similar as when assuming oracle source activity knowledge:
orthogonal additional vectors outperform random additional vectors,
except for lower \glspl{SNR} in dual-source segments,
and noise whitening outperforms noise subtraction and no noise handling,
with larger benefits for lower \glspl{SNR}.
In conclusion, the proposed \gls{BOPO-W} method combining
orthogonal additional vectors and noise whitening
also provides the best performance of all considered methods
when using online source counting.

\section{Conclusion}
\label{sec:conclusion}
In this paper, we proposed three extensions to the conventional \gls{BOP} method
to improve successive multi-source \gls{RTF} vector estimation.
First, we derived a closed-form solution
by setting the vector gradient of the \gls{BOP} cost function equal to zero,
significantly reducing the computational complexity of the \gls{BOP} method
compared to using an iterative gradient descent method.
Second, we provided a deeper understanding of the need and impact
of additional vectors in the optimization process,
and proposed to use orthogonal additional vectors
instead of random additional vectors in order to 
reduce undesired amplification of error signal components caused by model mismatch.
Third, we incorporated two noise handling techniques,
namely noise subtraction and noise whitening,
assuming an estimate of the noise covariance matrix is available.
Although not the primary focus of this paper, we also proposed
a frame-by-frame source counting method
based on the generalized magnitude-squared coherence
of the noisy covariance matrix whitened with the noisy covariance matrix
from some frames earlier.
Both when assuming oracle source activity knowledge as well as
when using the online source counting method,
simulation results in noisy reverberant scenarios demonstrate
that the proposed \gls{BOPO-W} method,
which combines the closed-form solution, orthogonal additional vectors, and noise whitening,
results in substantial improvements in terms of computational complexity,
\gls{RTF} vector estimation accuracy
and \gls{SINR} improvement compared to the conventional \gls{BOP} method.
These results demonstrate the robustness and practicality of our proposed extensions
for multi-source \gls{RTF} vector estimation
when sources activate successively.
Future work will focus on eliminating the need for additional vectors,
developing more sophisticated source counting methods
and handling source deactivation.


\appendices
\crefalias{section}{appsec} 
\section{Derivation of vector gradient}
\label{apx:vector_gradient_BOP}
The vector gradient $\nabla_{\bm{\theta}}J_{\mathbf{G}_{\bar{K}}}\!\!(\bm{\theta})$
of the \gls{BOP} cost function in~\Cref{eq:projected_costfun_noiseless}
can be derived using complex-valued matrix differentiation~\cite{hjorungnes2011complex}.
The gradient of a scalar real-valued function is given by 
$
    \nabla_{\bm{\theta}}J_{\mathbf{G}_{\bar{K}}}\!\!(\bm{\theta})
    = (\mathcal{D}_{\bm{\theta}}J_{\mathbf{G}_{\bar{K}}}\!\!(\bm{\theta}))^{\mathrm{H}}
$,
where $\mathcal{D}_{\bm{\theta}}$ denotes the complex-valued matrix derivative operator
with respect to the vector variable $\bm{\theta}$.
Inserting the \gls{BOP} cost function in~\Cref{eq:projected_costfun_noiseless},
applying the chain rule described in~\cite[Sec. 3.4.1]{hjorungnes2011complex}
and using~\cite[Example 4.13]{hjorungnes2011complex} yields
\begin{multline}
    \mathcal{D}_{\bm{\theta}}J_{\mathbf{G}_{\bar{K}}}\!\!\left(\bm{\theta}\right)
    = \mathcal{D}_{\bm{\theta}}\mathrm{tr}\{\mathbf{P}^{\angle}_{\mathbf{G}_{\bar{K}}\bm{\theta}}\mathbf{R}_{x,t}\mathbf{P}^{\angle\mathrm{H}}_{\mathbf{G}_{\bar{K}}\bm{\theta}}\} \\
    = \mathrm{vec}^{\mathrm{T}}\left\{\mathbf{I}_{M}\right\}\mathcal{D}_{\bm{\theta}}\left(\mathbf{P}^{\angle}_{\mathbf{G}_{\bar{K}}\bm{\theta}}\mathbf{R}_{x,t}\mathbf{P}^{\angle\mathrm{H}}_{\mathbf{G}_{\bar{K}}\bm{\theta}}\right),
    \label{eq:D_costfun}
\end{multline}
where $\mathrm{vec}\{\cdot\}$ denotes the vectorization operator,
which stacks the columns of a matrix into a column vector.
Applying the product rule from~\cite[Lemma 3.4]{hjorungnes2011complex} to~\Cref{eq:D_costfun} leads to
\begin{multline}
    \!\!\!\mathcal{D}_{\bm{\theta}}\!\left(\mathbf{P}^{\angle}_{\mathbf{G}_{\bar{K}}\bm{\theta}}\mathbf{R}_{x,t}\mathbf{P}^{\angle\mathrm{H}}_{\mathbf{G}_{\bar{K}}\bm{\theta}}\right)\!
    = \!\left(\!\left(\mathbf{R}_{x,t}\mathbf{P}^{\angle\mathrm{H}}_{\mathbf{G}_{\bar{K}}\bm{\theta}}\right)^{\!\mathrm{T}}\!\!\otimes\mathbf{I}_{M}\!\right)\!\mathcal{D}_{\bm{\theta}}\mathbf{P}^{\angle}_{\mathbf{G}_{\bar{K}}\bm{\theta}}\\
    + \left(\mathbf{I}_{M}\otimes\left(\mathbf{P}^{\angle}_{\mathbf{G}_{\bar{K}}\bm{\theta}}\mathbf{R}_{x,t}\right)\right)\mathcal{D}_{\bm{\theta}}\mathbf{P}^{\angle\mathrm{H}}_{\mathbf{G}_{\bar{K}}\bm{\theta}},
    \label{eq:D_matrixcostfun}
\end{multline}
where $\otimes$ denotes the Kronecker product.
Using the chain rule, identities involving complex conjugation~\cite[Lemma 3.3]{hjorungnes2011complex}
and results from~\cite[Table 4.4]{hjorungnes2011complex},
the derivative of the Hermitian oblique projection operator
$\mathcal{D}_{\bm{\theta}}\mathbf{P}^{\angle\mathrm{H}}_{\mathbf{G}_{\bar{K}}\bm{\theta}}$
in~\Cref{eq:D_matrixcostfun} is given by
\begin{align}
    \mathcal{D}_{\bm{\theta}}\mathbf{P}^{\angle\mathrm{H}}_{\mathbf{G}_{\bar{K}}\bm{\theta}} = \mathbf{C}_{M,M}\left(\mathcal{D}_{\bm{\theta}^{*}}\mathbf{P}^{\angle}_{\mathbf{G}_{\bar{K}}\bm{\theta}}\right)^{*},
    \label{eq:D_OPO_hermitian}
\end{align}
where $\mathbf{C}_{M,N}\in\{0,1\}^{MN\times MN}$ denotes the commutation matrix
of an $M\times N$ dimensional matrix $\mathbf{A}$,
i.e., $\mathbf{C}_{M,N}\mathrm{vec}\{\mathbf{A}\}
= \mathrm{vec}\{\mathbf{A}^{\mathrm{T}}\}$.
The derivative of the oblique projection operator
$\mathcal{D}_{\bm{\theta}}\mathbf{P}^{\angle}_{\mathbf{G}_{\bar{K}}\bm{\theta}}$
in~\Cref{eq:D_matrixcostfun}
can be derived using the product rule, i.e.,
\begin{multline}
    \mathcal{D}_{\bm{\theta}}\mathbf{P}^{\angle}_{\mathbf{G}_{\bar{K}}\bm{\theta}}
    = \mathcal{D}_{\bm{\theta}}\left(\mathbf{G}_{\bar{K}}\left(\mathbf{G}_{\bar{K}}^{\mathrm{H}}\mathbf{P}^{\perp}_{\bm{\theta}}\mathbf{G}_{\bar{K}}\right)^{-1}\mathbf{G}_{\bar{K}}^{\mathrm{H}}\mathbf{P}^{\perp}_{\bm{\theta}}\right) \\
    = \left(\left(\mathbf{G}_{\bar{K}}^{\mathrm{H}}\mathbf{P}^{\perp}_{\bm{\theta}}\right)^{\mathrm{T}}\otimes\mathbf{G}_{\bar{K}}\right)\mathcal{D}_{\bm{\theta}}\left(\mathbf{G}_{\bar{K}}^{\mathrm{H}}\mathbf{P}^{\perp}_{\bm{\theta}}\mathbf{G}_{\bar{K}}\right)^{-1} \\ + \left(\mathbf{I}_{M}\otimes\mathbf{G}_{\bar{K}}\left(\mathbf{G}_{\bar{K}}^{\mathrm{H}}\mathbf{P}^{\perp}_{\bm{\theta}}\mathbf{G}_{\bar{K}}\right)^{-1}\mathbf{G}_{\bar{K}}^{\mathrm{H}}\right) \mathcal{D}_{\bm{\theta}}\mathbf{P}^{\perp}_{\bm{\theta}}.
    \label{eq:D_OPO}
\end{multline}
Using the chain rule, the derivative of the matrix inversion~\cite[Example 4.23, Table 4.4]{hjorungnes2011complex} and the product rule leads to
\begin{multline}
    \!\!\!\!\!\mathcal{D}_{\bm{\theta}}\left(\mathbf{G}_{\bar{K}}^{\mathrm{H}}\mathbf{P}^{\perp}_{\bm{\theta}}\mathbf{G}_{\bar{K}}\right)^{-1} \\
    \!\!\!\!=\!-\!\left(\!\left(\mathbf{G}_{\bar{K}}^{\mathrm{H}}\mathbf{P}^{\perp}_{\bm{\theta}}\mathbf{G}_{\bar{K}}\right)^{\!-\mathrm{T}}\!\!\!\otimes\!\left(\mathbf{G}_{\bar{K}}^{\mathrm{H}}\mathbf{P}^{\perp}_{\bm{\theta}}\mathbf{G}_{\bar{K}}\right)^{\!-1}\right)\!\mathcal{D}_{\bm{\theta}}\!\left(\mathbf{G}_{\bar{K}}^{\mathrm{H}}\mathbf{P}^{\perp}_{\bm{\theta}}\mathbf{G}_{\bar{K}}\right) \\ \!\!\!=\!-\!\left(\!\!\left(\!\mathbf{G}_{\bar{K}}\!\left(\mathbf{G}_{\bar{K}}^{\mathrm{H}}\mathbf{P}^{\perp}_{\bm{\theta}}\mathbf{G}_{\bar{K}}\right)^{\!\!-1}\right)^{\!\!\mathrm{T}}\!\!\!\!\otimes\!\!\Big(\!\mathbf{G}_{\bar{K}}^{\mathrm{H}}\mathbf{P}^{\perp}_{\bm{\theta}}\mathbf{G}_{\bar{K}}\!\Big)^{\!\!-1}\!\!\!\!\!\mathbf{G}_{\bar{K}}^{\mathrm{H}}\!\!\right)\!\mathcal{D}_{\bm{\theta}}\mathbf{P}^{\perp}_{\bm{\theta}}\!.
    \label{eq:D_inverse_within_OPO}
\end{multline}
Substituting~\Cref{eq:D_inverse_within_OPO} into~\Cref{eq:D_OPO},
using the Kronecker product identity from~\cite[Lemma 2.10]{hjorungnes2011complex}
and merging terms yields
\begin{align}
    \!\!\!\!\mathcal{D}_{\bm{\theta}}\mathbf{P}^{\angle}_{\mathbf{G}_{\bar{K}}\bm{\theta}}  \!\!
    =\!\! \left(\!\!\!\left(\!\mathbf{I}_{M}\!\!-\!\!\mathbf{P}^{\angle}_{\mathbf{G}_{\bar{K}}\bm{\theta}}\!\right)^{\!\!\mathrm{T}}\!\!\!\!\otimes\!\mathbf{G}_{\bar{K}}\!\Big(\!\mathbf{G}_{\bar{K}}^{\mathrm{H}}\mathbf{P}^{\perp}_{\bm{\theta}}\mathbf{G}_{\bar{K}}\!\Big)^{\!\!-1}\!\!\!\!\!\mathbf{G}_{\bar{K}}^{\mathrm{H}}\!\!\right)\!\mathcal{D}_{\bm{\theta}}\mathbf{P}^{\perp}_{\bm{\theta}}\!.
    \label{eq:D_oblique_depending_on_orthogonal}
\end{align}
Note that no differentiation in~\Crefrange{eq:D_costfun}{eq:D_oblique_depending_on_orthogonal}
depends directly on $\bm{\theta}$, so that
$\mathcal{D}_{\bm{\theta}^{*}}\mathbf{P}^{\angle}_{\mathbf{G}_{\bar{K}}\bm{\theta}}$ in~\Cref{eq:D_OPO_hermitian}
can be derived by exchanging $\mathcal{D}_{\bm{\theta}}$ with $\mathcal{D}_{\bm{\theta}^{*}}$
in~\Cref{eq:D_oblique_depending_on_orthogonal}, i.e.,
\begin{align}
    \mathcal{D}_{\bm{\theta}^{*}}\mathbf{P}^{\angle}_{\mathbf{G}_{\bar{K}}\bm{\theta}}  \!
    =\!\! \left(\!\!\left(\!\mathbf{I}_{M}\!-\!\mathbf{P}^{\angle}_{\mathbf{G}_{\bar{K}}\bm{\theta}}\!\right)^{\!\mathrm{T}}\!\!\!\!\!\otimes\!\mathbf{G}_{\bar{K}}\!\Big(\!\mathbf{G}_{\bar{K}}^{\mathrm{H}}\mathbf{P}^{\perp}_{\bm{\theta}}\mathbf{G}_{\bar{K}}\!\Big)^{\!-1}\!\!\!\!\mathbf{G}_{\bar{K}}^{\mathrm{H}}\!\right)\!\mathcal{D}_{\bm{\theta}^{*}}\mathbf{P}^{\perp}_{\bm{\theta}}\!.
    \label{eq:Dconj_oblique_depending_on_orthogonal}
\end{align}
The derivative of the complement orthogonal projection operator
$\mathcal{D}_{\bm{\theta}}\mathbf{P}^{\perp}_{\bm{\theta}}$ in~\Cref{eq:D_oblique_depending_on_orthogonal}
can be derived
by first applying the product rule, i.e.,
\begin{multline}
    \!\!\!\!\!\mathcal{D}_{\bm{\theta}}\mathbf{P}^{\perp}_{\bm{\theta}}
    = \mathcal{D}_{\bm{\theta}}\!\left(\mathbf{I}_{M}-\bm{\theta}\left(\bm{\theta}^{\mathrm{H}}\bm{\theta}\right)^{-1}\!\bm{\theta}^{\mathrm{H}}\right)
    =  - \left(\bm{\theta}^{+\mathrm{T}}\!\otimes\mathbf{I}_{M}\right)\mathcal{D}_{\bm{\theta}}\bm{\theta} \\
    - \left(\mathbf{I}_{M}\otimes\bm{\theta}^{+\mathrm{H}}\right)\mathcal{D}_{\bm{\theta}}\bm{\theta}^{\mathrm{H}}-\left(\bm{\theta}^{*}\otimes\bm{\theta}\right)\mathcal{D}_{\bm{\theta}}\left(\bm{\theta}^{\mathrm{H}}\bm{\theta}\right)^{-1}.
    \label{eq:D_orthogonal_projection}
\end{multline}
Using the chain rule, the derivative of the matrix inversion in~\Cref{eq:D_orthogonal_projection} is given by
\begin{align}
     & \mathcal{D}_{\bm{\theta}}\left(\bm{\theta}^{\mathrm{H}}\bm{\theta}\right)^{-1}
    = -\left(\left(\bm{\theta}^{\mathrm{H}}\bm{\theta}\right)^{-1}\otimes\left(\bm{\theta}^{\mathrm{H}}\bm{\theta}\right)^{-1}\right)\mathcal{D}_{\bm{\theta}}\left(\bm{\theta}^{\mathrm{H}}\bm{\theta}\right),
    \label{eq:D_thetathetaH_inverse}
    \\
     & \text{with}\quad
    \mathcal{D}_{\bm{\theta}}\left(\bm{\theta}^{\mathrm{H}}\bm{\theta}\right)
    = \bm{\theta}^{\mathrm{T}}\mathcal{D}_{\bm{\theta}}\bm{\theta}^{\mathrm{H}}
    + \bm{\theta}^{\mathrm{H}}\mathcal{D}_{\bm{\theta}}\bm{\theta},
    \label{eq:D_thetathetaH}
\end{align}
using the product rule.
Inserting~\Cref{eq:D_thetathetaH_inverse,eq:D_thetathetaH} into~\Cref{eq:D_orthogonal_projection}
and evaluating $\mathcal{D}_{\bm{\theta}}\bm{\theta}^{\mathrm{H}}$ and $\mathcal{D}_{\bm{\theta}}\bm{\theta}$ according to~\cite[Table 4.4]{hjorungnes2011complex} yields
\begin{align}
    \mathcal{D}_{\bm{\theta}}\mathbf{P}^{\perp}_{\bm{\theta}} = 
    - \left(\bm{\theta}^{+\mathrm{T}}\otimes\mathbf{P}^{\perp}_{\bm{\theta}}\right).
    \label{eq:D_orthogonal_proj_dep_on_D_theta}
\end{align}
Using  identities involving complex conjugation,
$\mathbf{P}^{\perp}_{\bm{\theta}}=\mathbf{P}^{\perp\mathrm{H}}_{\bm{\theta}} = (\mathbf{P}^{\perp\mathrm{T}}_{\bm{\theta}})^{*}$,
the chain rule, \Cref{eq:D_orthogonal_proj_dep_on_D_theta},
and the Kronecker product in combination
with commutation matrices~\cite[Lemma 2.12]{hjorungnes2011complex},
the derivative
$\mathcal{D}_{\bm{\theta}^{*}}\mathbf{P}^{\perp}_{\bm{\theta}}$ in~\Cref{eq:Dconj_oblique_depending_on_orthogonal}
can be derived as
\begin{align}
    \!\!\!\!\mathcal{D}_{\bm{\theta}^{*}}\mathbf{P}^{\perp}_{\bm{\theta}} \!\!
    =\!\! \left(\mathcal{D}_{\bm{\theta}}\mathbf{P}^{\perp\mathrm{T}}_{\bm{\theta}}\right)^{\!*} \!\!\!
    =\!\! \left(\mathbf{C}_{M,M}\mathcal{D}_{\bm{\theta}}\mathbf{P}^{\perp}_{\bm{\theta}}\right)^{\!*} \!\!\!
    =\! -\! \left(\mathbf{P}^{\perp\mathrm{T}}_{\bm{\theta}}\!\otimes\!\bm{\theta}^{+\mathrm{H}}\right)\!.
    \label{eq:Dconj_orthogonal_proj_dep_on_D_theta}
\end{align}
Inserting~\Cref{eq:D_orthogonal_proj_dep_on_D_theta}
into~\Cref{eq:D_oblique_depending_on_orthogonal} 
and~\Cref{eq:Dconj_orthogonal_proj_dep_on_D_theta}
into~\Cref{eq:Dconj_oblique_depending_on_orthogonal},
and using the Kronecker product identity
from~\cite[Lemma 2.10]{hjorungnes2011complex}
and~\Cref{eq:oblique_projection} yields
\begin{align}
    \mathcal{D}_{\bm{\theta}}\mathbf{P}^{\angle}_{\mathbf{G}_{\bar{K}}\bm{\theta}}
    =& \left(\bm{\theta}^{+}\left(\mathbf{P}^{\angle}_{\mathbf{G}_{\bar{K}}\bm{\theta}}-\mathbf{I}_{M}\right)\right)^{\mathrm{T}}\otimes\mathbf{P}^{\angle}_{\mathbf{G}_{\bar{K}}\bm{\theta}}.
    \label{eq:D_OPO_full}
    \\
    \mathcal{D}_{\bm{\theta}^{*}}\mathbf{P}^{\angle}_{\mathbf{G}_{\bar{K}}\bm{\theta}}  \!
    =&\!\! \left(\!\mathbf{P}^{\perp}_{\bm{\theta}}\!\left(\!\mathbf{P}^{\angle}_{\mathbf{G}_{\bar{K}}\bm{\theta}}\!-\!\mathbf{I}_{M}\!\right)\!\right)^{\!\mathrm{T}}\!\!\!\!\otimes\!\mathbf{G}_{\bar{K}}\!\Big(\!\mathbf{G}_{\bar{K}}^{\mathrm{H}}\mathbf{P}^{\perp}_{\bm{\theta}}\mathbf{G}_{\bar{K}}\!\Big)^{\!-1}\!\!\!\!\mathbf{G}_{\bar{K}}^{\mathrm{H}}\bm{\theta}^{+\mathrm{H}}\!.
    \label{eq:Dconj_OPO}
\end{align}
Using $\!\mathbf{I}_{M}\!\!-\!\!\mathbf{P}^{\angle}_{\mathbf{G}_{\bar{K}}\bm{\theta}} \!\!=\!\! \mathbf{P}^{\angle}_{\bm{\theta}\mathbf{G}_{\bar{K}}}\!\!\!+\!\!\mathbf{P}^{\perp}_{[\mathbf{G}_{\bar{K}},\bm{\theta}]}\!$
simplifies~\Cref{eq:D_OPO_full,eq:Dconj_OPO} to
\begin{align}
    \mathcal{D}_{\bm{\theta}}\mathbf{P}^{\angle}_{\mathbf{G}_{\bar{K}}\bm{\theta}}  \!
    = & \!-\!\left(\bm{\theta}^{+}\mathbf{P}^{\angle}_{\bm{\theta}\mathbf{G}_{\bar{K}}}\right)^{\mathrm{T}}\otimes\mathbf{P}^{\angle}_{\mathbf{G}_{\bar{K}}\bm{\theta}}\label{eq:D_OPO_full_simplified}                                                                              \\
    \mathcal{D}_{\bm{\theta}^{*}}\mathbf{P}^{\angle}_{\mathbf{G}_{\bar{K}}\bm{\theta}}  \!
    = & \!-\!\mathbf{P}^{\perp\mathrm{T}}_{[\mathbf{G}_{\bar{K}},\bm{\theta}]}\otimes\mathbf{G}_{\bar{K}}\Big(\!\mathbf{G}_{\bar{K}}^{\mathrm{H}}\mathbf{P}^{\perp}_{\bm{\theta}}\mathbf{G}_{\bar{K}}\!\Big)^{\!-1}\!\!\!\mathbf{G}_{\bar{K}}^{\mathrm{H}}\bm{\theta}^{+\mathrm{H}}.
    \label{eq:Dconj_full_simplified}
\end{align}
Inserting~\Cref{eq:Dconj_full_simplified} into~\Cref{eq:D_OPO_hermitian} and using~\cite[Lemma 2.12]{hjorungnes2011complex} yields
\begin{align}
    \mathcal{D}_{\bm{\theta}}\mathbf{P}^{\angle\mathrm{H}}_{\mathbf{G}_{\bar{K}}\bm{\theta}} \!
    =\! -\!\left(\!\bm{\theta}^{+}\mathbf{G}_{\bar{K}}\Big(\!\mathbf{G}_{\bar{K}}^{\mathrm{H}}\mathbf{P}^{\perp}_{\bm{\theta}}\mathbf{G}_{\bar{K}}\!\Big)^{\!-1}\!\!\!\mathbf{G}_{\bar{K}}^{\mathrm{H}}\right)^{\!\mathrm{T}}\!\!\!\!\otimes\!\mathbf{P}^{\perp}_{[\mathbf{G}_{\bar{K}},\bm{\theta}]}.
    \label{eq:D_OPO_hermitian_full}
\end{align}
Inserting~\Cref{eq:D_matrixcostfun,eq:D_OPO_full_simplified,eq:D_OPO_hermitian_full} into~\Cref{eq:D_costfun} yields
\begin{multline}
    \mathcal{D}_{\bm{\theta}}J_{\mathbf{G}_{\bar{K}}}\!\!\left(\bm{\theta}\right) \!
    =\!-\mathrm{vec}^{\mathrm{T}}\!\left\{\mathbf{I}_{M}\right\}\!   \bigg(\!\!\!\left(\!\bm{\theta}^{+}\mathbf{P}^{\angle}_{\bm{\theta}\mathbf{G}_{\bar{K}}}\mathbf{R}_{x,t}\mathbf{P}^{\angle\mathrm{H}}_{\mathbf{G}_{\bar{K}}\bm{\theta}}\!\right)^{\!\mathrm{T}}\!\!\!\!\otimes\mathbf{P}^{\angle}_{\mathbf{G}_{\bar{K}}\bm{\theta}} \\ +\! \left(\!\bm{\theta}^{+}\mathbf{G}_{\bar{K}}\Big(\!\mathbf{G}_{\bar{K}}^{\mathrm{H}}\mathbf{P}^{\perp}_{\bm{\theta}}\mathbf{G}_{\bar{K}}\!\Big)^{\!-1}\!\!\!\!\mathbf{G}_{\bar{K}}^{\mathrm{H}}\!\right)^{\!\mathrm{T}}\!\!\!\!\otimes\mathbf{P}^{\angle}_{\mathbf{G}_{\bar{K}}\bm{\theta}}\mathbf{R}_{x,t}\mathbf{P}^{\perp}_{[\mathbf{G}_{\bar{K}},\bm{\theta}]}\!\bigg).
    \label{eq:D_final_nasty}
\end{multline}
Using the relation between the Kronecker product
and the vectorization operator in~\cite[Lemma 2.11]{hjorungnes2011complex},
\Cref{eq:D_final_nasty} simplifies to
\begin{multline}
    \mathcal{D}_{\bm{\theta}}J_{\mathbf{G}_{\bar{K}}}\!\!\left(\bm{\theta}\right) = 
    -\bm{\theta}^{+}\mathbf{P}^{\angle}_{\bm{\theta}\mathbf{G}_{\bar{K}}}\mathbf{R}_{x,t}\mathbf{P}^{\angle\mathrm{H}}_{\mathbf{G}_{\bar{K}}\bm{\theta}}\mathbf{P}^{\angle}_{\mathbf{G}_{\bar{K}}\bm{\theta}} \\
    - \bm{\theta}^{+}\mathbf{G}_{\bar{K}}\left(\mathbf{G}_{\bar{K}}^{\mathrm{H}}\mathbf{P}^{\perp}_{\bm{\theta}}\mathbf{G}_{\bar{K}}\right)^{-1}\mathbf{G}_{\bar{K}}^{\mathrm{H}}\mathbf{P}^{\angle}_{\mathbf{G}_{\bar{K}}\bm{\theta}}\mathbf{R}_{x,t}\mathbf{P}^{\perp}_{[\mathbf{G}_{\bar{K}},\bm{\theta}]}.
    \label{eq:D_final}
\end{multline}
Inserting~\Cref{eq:D_final} into
$
    \nabla_{\bm{\theta}}J_{\mathbf{G}_{\bar{K}}}\!\!(\bm{\theta})
    = (\mathcal{D}_{\bm{\theta}}J_{\mathbf{G}_{\bar{K}}}\!\!(\bm{\theta}))^{\mathrm{H}}
$
yields the vector gradient of the cost function given in~\Cref{eq:BOPgradient}.

\bibliographystyle{IEEEtran}
\bibliography{mybib}







\vfill

\end{document}